\documentclass{aastex63}

\shorttitle{homologous eruptions}
\shortauthors{Sahu et al.}

\begin{document}
{


\title{Evolution of magnetic fields and energy release processes during homologous eruptive flares}

\correspondingauthor{Suraj Sahu}
\email{suraj@prl.res.in, sahusuraj419@gmail.com}

\author{Suraj Sahu}
\affiliation{Udaipur Solar Observatory, Physical Research Laboratory, Dewali, Badi Road, Udaipur-313 001, Rajasthan, India}
\affiliation{Discipline of Physics, Indian Institute of Technology  Gandhinagar, Palaj, Gandhinagar-382 355, Gujarat, India}

\author{Bhuwan Joshi}
\affiliation{Udaipur Solar Observatory, Physical Research Laboratory, Dewali, Badi Road, Udaipur-313 001, Rajasthan, India}

\author{Avijeet Prasad}
\affiliation{Rosseland Centre for Solar Physics, University of Oslo, Postboks 1029, Blindern NO-0315, Oslo, Norway}

\affiliation{Institute of Theoretical Astrophysics, University of Oslo, Postboks 1029, Blindern NO-0315, Oslo, Norway}



\author{Kyung-Suk Cho}
\affiliation{Space Science Division, Korea Astronomy and Space Science Institute, Daejeon 305-348, Republic of Korea}

\affiliation{Department of Astronomy and Space Science, University of Science and Technology, Daejeon 305-348, Republic of Korea}

\begin{abstract}

We explore the processes of repetitive build-up and explosive release of magnetic energy together with the formation of magnetic flux ropes that eventually resulted into three homologous eruptive flares of successively increasing intensities (i.e., M2.0, M2.6, and X1.0).
The flares originated from NOAA active region 12017 during 2014 March 28--29. EUV observations and magnetogram measurements together with coronal magnetic field modeling suggest that the flares were triggered by the eruption of flux ropes embedded by a densely packed system of loops within a small part of the active region. In X-rays, the first and second events show similar evolution with single, compact sources, while the third event exhibits multiple emission centroids with a set of strong non-thermal conjugate sources at 50--100 keV during the HXR peak. 
The photospheric magnetic field over an interval of $\approx$44 hr encompassing the three flares undergoes important phases of emergence and cancellation processes together with significant changes near the polarity inversion lines within the flaring region. Our observations point toward the tether-cutting mechanism as the plausible triggering process of the eruptions.
Between the second and third event, we observe a prominent phase of flux emergence which temporally correlates with the build-up phase of free magnetic energy in the active region corona. 
In conclusion, our analysis reveals an efficient coupling between the rapidly evolving photospheric and coronal magnetic fields in the active region that led to a continued phase of the build-up of free energy, resulting into the homologous flares of successively increasing intensities.

\end{abstract}


\section{Introduction} \label{sec:intro}

Solar flares are the sudden, explosive events in the solar atmosphere, which release huge amount of energy in the form of heat, radiation, bulk plasma motion, and produce highly accelerated charged particles \citep{Fletcher2011,Benz2017}. It is widely believed that the fundamental processes that drive an eruptive event--build-up/storage of free magnetic energy and its explosive release via magnetic reconnection--are inherently guided by the complexity of solar magnetic field \citep{Priest2002}.
Therefore, it is important to analyze the variability in the build-up and release of magnetic energy to understand the driver of solar flares and associated processes. The solar eruptive phenomena involve the expulsion of magnetized plasma out into the heliosphere. Hence, in order to understand the dynamics of magnetized plasma during and subsequent to the explosive energy release, we need to explore the multi-wavelength and multi-instrument data together with coronal magnetic field modeling.

The strongest magnetic field regions of the Sun are known as solar active regions. The active regions present a diverse nature in their morphology depending upon the distribution and strength of the underlying photospheric magnetic fields \citep{Toriumi2019}. Typically, during its growth phase, as an active region expands and evolves, the complexity in the photospheric magnetic field increases.
A complex active region may produce several energetic events (viz., flares, coronal mass ejections, jets, prominence eruptions, etc.) during its whole lifetime \citep{Joshi2018,Mitra2018,Mitra2020a,Sahu2020,Zuccarello2021}.

Solar eruptions in a repetitive manner may originate from the same location of the active region and sometimes they show morphological resemblance in the multi-wavelength imaging and coronagraphic observations. Such repetitive activities are known as `homologous eruptions' \citep{Woodgate1984,Zhang2002}. The exploration of homologous eruptions is extremely important to understand the role of photospheric magnetic field variations and associated coronal changes in determining the eruptivity. In this way, by assessing the homology tendency of an active region, we can provide important inputs toward the understanding of onset of coronal mass ejections (CMEs) and subsequent space-weather consequences. In the past, several studies on different features of homologous eruptions were carried out that reveal following aspects to be responsible for the occurrence of homologous activity: flux emergence \citep{Nitta2001,Chatterjee2013}, shearing motion and magnetic flux cancellation \citep{Li2010,Vemareddy2017}, persistent photospheric horizontal motion of the magnetic structure along the polarity inversion line (PIL) \citep{Romano2015,Romano2018}, coronal null point configuration \citep{Devore2008}, etc. The homologous solar eruptions form a contemporary research topic in solar physics and the present study aims to provide additional observational inputs in this direction.

In this study, we investigate a detailed evolution of photospheric magnetic fields associated with the three homologous eruptive flares occurred during 2014 March 28--29 in the NOAA active region (AR) 12017. Interestingly, the three homologous events are of successively increasing intensities (M2.0, M2.6, X1.0). In our previous study \citep{Sahu2022}, we have explored the formation process of three homologous, broad CMEs resulting from these three eruptive flares. 
We have identified the events as flux rope eruptions originating from the same compact flaring region of the AR. Our work have presented a clear example of a large-scale coronal magnetic configuration that is repeatedly blown out by compact flux rope eruptions leading to a series of broad CMEs.
Flaring activities in AR 12017, especially the X-class event on 2014 March 29, have been subjected to various studies pertaining to observational and modeling analyses \citep{Kleint2015,Li2015,Liu2015,Young2015,Yang2016,Woods2017,Woods2018,Cheung2019}. \citet{Liu2015} discussed a scenario of asymmetric filament eruption in the context of nonuniform filament confinement and an MHD instability prior to the X-flare. The study by \citet{Yang2016} was comprised of all the flaring activities in the AR during 2014 March 28--29. They concluded that the flares were triggered mainly by the kink instability of the associated filaments. \citet{Woods2018} investigated the triggering mechanism of the third event (X1.0 flare) and associated filament eruption. Their study confirmed the existence of two flux ropes within the active region prior to flaring. Interestingly, one of these two flux ropes erupts, which might be due to the tether cutting reconnection \citep{Moore2001} allowing the flux rope to rise to a torus unstable region.
In this paper, our motivation is to conduct a detailed study to understand the repetitive build-up of magnetic energy and flux ropes that eventually drive the three homologous eruptive flares. Toward this, we provide a quantitative estimation of the temporal evolution of free magnetic energy in the AR and examine precisely the changes in the photospheric magnetic flux during the entire period of homologous flaring activity. We further present a detailed multi-wavelength investigation of the temporal, spatial, and spectral characteristics of each event.
In Section \ref{sec:observation_data}, we provide a brief discussion of the data sources and analysis techniques. Section \ref{sec:multiwavelength} gives the details of extreme ultraviolet (EUV) and X-ray observations of the flares. In Section \ref{sec:magnetic_evolution}, we describe the photospheric magnetic field evolution during the events and associated coronal magnetic configuration. The build-up of photospheric current in relation to the triggering of the eruptions is presented in Section \ref{sec:current}. The evolutionary stages of the eruptive hot plasma structures are presented in Section \ref{sec:time_slice}. The details of the storage and release processes of free magnetic energy are described in Section \ref{sec:free_energy}. We discuss and interpret our results in the final section.

\section{OBSERVATIONAL DATA SOURCES AND TECHNIQUES}
\label{sec:observation_data}

We use data from the Atmospheric Imaging Assembly \citep[AIA;][]{Lemen2012} on board the Solar Dynamics Observatory \citep[SDO;][]{Pesnell2012} for EUV imaging and analysis. The AIA observes the full disk of the Sun in seven EUV (94 \AA, 131 \AA, 171 \AA, 193 \AA, 211 \AA, 304 \AA, and 335 \AA), two ultraviolet (UV) (1600 \AA\ and 1700 \AA), and one visible (4500 \AA) channels. For our analysis, we use the EUV 304 \AA\ [log(T) $\approx$4.7] and 193 \AA\ [log(T) $\approx$6.2, 7.3] observations. The 304 \AA\ images provide information of the chromosphere and transition region of the Sun, while the 193 \AA\ images are used to analyze the corona and hot flare plasma. In order to investigate the evolution of the photosphere through the line-of-sight (LOS) magnetogram and continuum observations, we obtain data from the Helioseismic Magnetic Imager \citep[HMI;][]{Schou2012} on board the SDO. 

To visualize the X-ray sources and to quantify the parameters associated with the X-ray emission processes, we use data obtained from the Reuven Ramaty High Energy Solar Spectroscopic Imager \citep[RHESSI;][]{Lin2002}. RHESSI observes the full-disk solar X-ray sources in the energy range of 3 keV to 17 MeV. We use the CLEAN algorithm \citep{Hurford2002} to reconstruct the X-ray images in different energy bands (i.e, 3--6, 6-12, 12--25, 25--50, and 50--100 keV). For image reconstruction, we use the front segments of the detectors 3--8 with 20 s integration time. We also carry out X-ray spectroscopy using the RHESSI data. The details of spectroscopy are given in Section \ref{sec:spectra}.

For the calculation of free magnetic energy associated with the AR in the coronal volume, we need the 3D magnetic field information above the photosphere. For this purpose, we use the nonlinear force-free field (NLFFF) extrapolation technique, which was originally formulated by \citet{Wiegelmann2008} and was further developed by \citet{Wiegelmann2010} and \citet{Wiegelmann2012}. We use the vector magnetograms (HMI.sharp$\textunderscore$cea$\textunderscore$720s series) as the photospheric input boundary conditions for the extrapolation. The extrapolation volume extends upto 280, 229, and 183 Mm in X, Y, and Z directions, respectively, considering the photosphere as the X-Y plane. 


\begin{deluxetable}{cccccc}[t!]
\tablecolumns{6}
\tablewidth{6pt}

 \tablecaption{The flares in NOAA AR 12017 during 2014 March 28--29 \label{tab:summary}}

 \tablehead{
 \colhead{\hspace{-0.1cm} Flares} & \colhead{\hspace{0.5cm}Flare} & \colhead{\hspace{0.5cm}Date} & \colhead{} & \colhead{Time (UT)} & \colhead{}\\
\colhead{\vspace{-0.1cm}} & \colhead{\hspace{0.5cm}class} & \colhead{} & \colhead{\hspace{0.5cm}Start} & \colhead{\hspace{0.5cm}Peak} & \colhead{\hspace{0.5cm}End}\
}

 \startdata 
 F1 & \hspace{0.5cm}M2.0 & \hspace{0.5cm}2014 March 28 & \hspace{0.5cm}19:05 & \hspace{0.5cm}19:18 & \hspace{0.5cm}19:27\\
F2 & \hspace{0.5cm}M2.6 & \hspace{0.5cm}2014 March 28 & \hspace{0.5cm}23:44 & \hspace{0.5cm}23:51 & \hspace{0.5cm}23:58\\
F3 & \hspace{0.5cm}X1.0 & \hspace{0.5cm}2014 March 29 & \hspace{0.5cm}17:35 & \hspace{0.5cm}17:48 & \hspace{0.5cm}17:54\\
 \enddata

\end{deluxetable}

\begin{figure}[t!]
\centering
\includegraphics[scale=1.5]{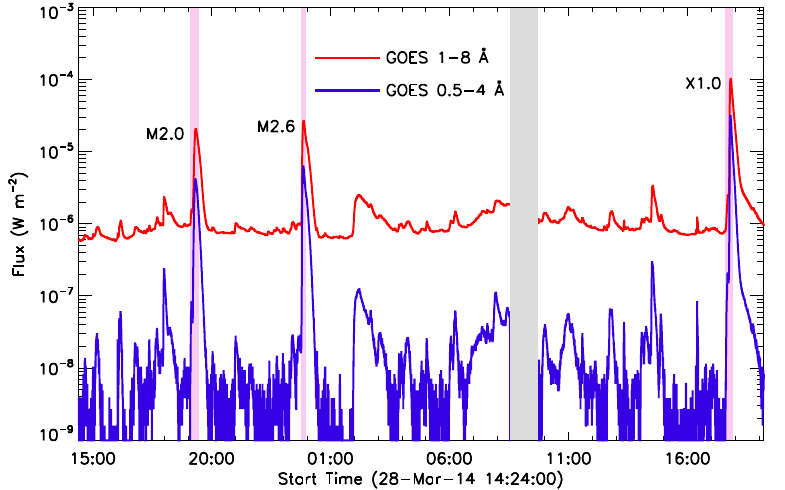}
\caption{GOES light curves in 1--8 and 0.5--4 \AA\ channels showing the three flares (M2.0, M2.6, and X1.0) under analysis. The intervals of the three flares are shown by pink vertical stripes. The gray shaded region indicates an interval when the GOES data were unavailable.}
\label{fig:goes}
\end{figure}

\section{MULTI-WAVELENGTH ANALYSIS OF FLARE EVOLUTION}
\label{sec:multiwavelength}

\subsection{Multi-wavelength overview of AR 12017}

Our study incorporates three homologous flaring events of successively increasing intensities (M2.0, M2.6, X1.0). The flares occurred in the NOAA AR 12017 during 2014 March 28--29. A summary of the flares is given in Table \ref{tab:summary}, which is based on the GOES
flare catalog\footnote{\url{https://www.ngdc.noaa.gov/stp/space-weather/solar-data/solar-features/solar-flares/x-rays/goes/xrs/goes-xrs-report_2014.txt}}. The three flares are indicated over the GOES light curves (in 1--8 and 0.5--4 \AA\ channels) in Figure \ref{fig:goes}. The durations of the three flares are marked by vertical pink stripes. The gray shaded region indicates an interval, when the GOES data were unavailable.

In Figure \ref{fig:overview}, we provide a multi-wavelength view of AR 12017 by plotting simultaneous white light continuum, magnetogram, and EUV images prior to the onset of the first event of M2.0 intensity. By comparing different panels of Figure \ref{fig:overview}, one can note many interesting features of the AR and flaring region. A comparison of continuum and magnetogram images (see panels (a) and (b)) suggests that the leading part of the AR consists of sunspots of predominantly negative polarity (see the region enclosed by the box in various panels of Figure \ref{fig:overview}), which happen to be the source of eruptive flares under analysis. Hence, we term the region inside the box as the `flaring region' (FR). On the trailing part of the AR, we observe sparsely located small sunspots with dispersed flux of predominantly positive polarity (see Figures \ref{fig:overview}(a)--(b)). The EUV images at 171 and 193 \AA\ readily suggest the existence of large coronal loops connecting the leading and trailing parts of the AR (Figures \ref{fig:overview}(c)--(d)). The presence of compact, closed loop configuration in the FR is also clearly visible. In Figure \ref{fig:overview}(e), we provide a preflare 304 \AA\ image of AR 12017. Here, we can clearly distinguish the dominance of brighter emission from the FR over other parts of the AR.

\begin{figure}
\centering
\includegraphics[scale=1.2]{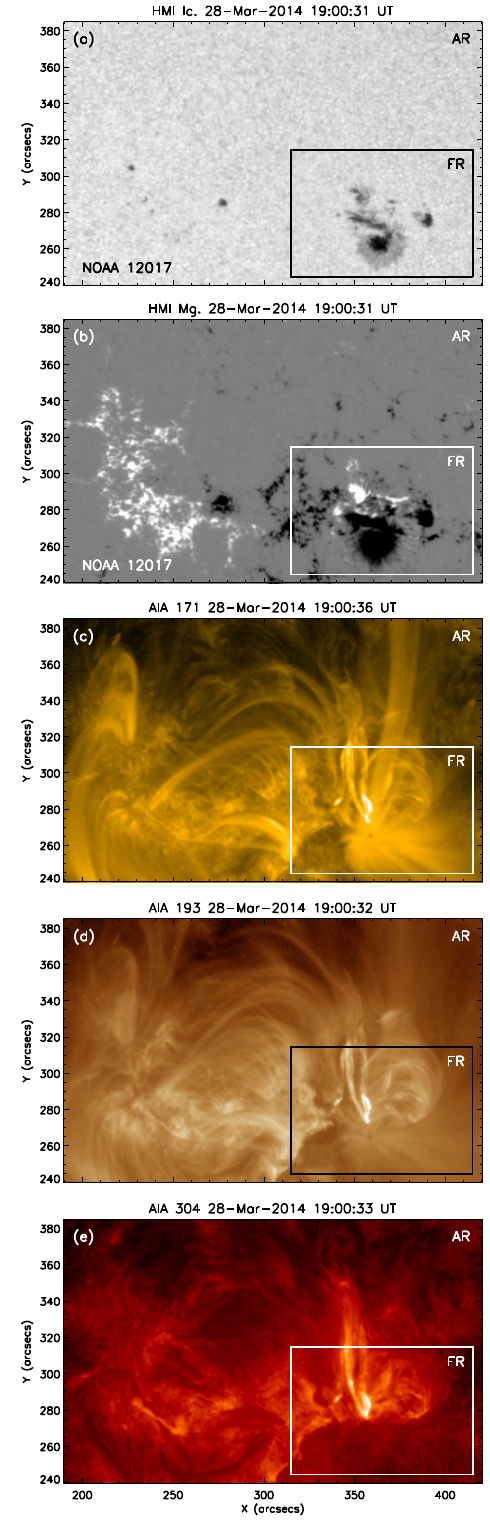}
\caption{Multi-wavelength view of AR 12017 on 2014 March 28 at $\approx$19:00 UT (i.e., 5 minutes before the start of the first flare according to GOES data). Panel (a): HMI continuum image of the AR. We observe a large sunspot in the leading part of the AR, which we mark by a black box. The flares of our study occurred within this marked area of the AR, which we term as the flaring region (FR). We indicate the FR by a box in all the subsequent panels. Panel (b): HMI LOS magnetogram of the AR. The FR consists of strong negative polarity region together with relatively weaker, compact positive polarities, located north of it. Panels (c) and (d): AIA 171 \AA\ [log(T) $\approx$5.7] and 193 \AA\ [log(T) $\approx$6.2, 7.3] images showing the connectivity of different loop systems between the leading and trailing parts of the AR along with compact loops within the FR. Panel (e): AIA 304 \AA\ [log(T) $\approx$4.7] image displaying much brighter emission from the FR, compared to other parts of the AR.}
\label{fig:overview}
\end{figure}

\begin{figure}
\centering
\includegraphics[scale=0.7]{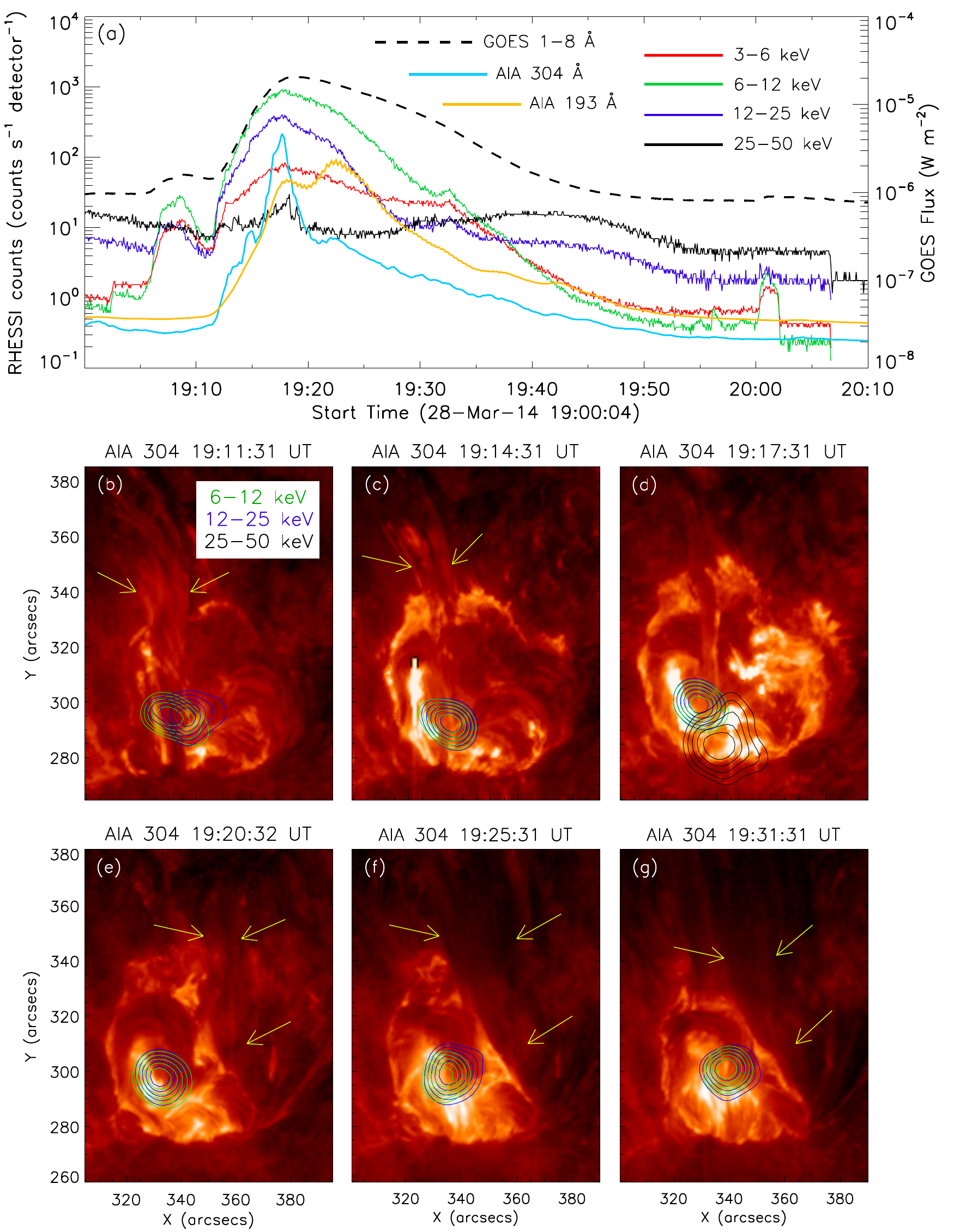}
\caption{Panel (a): RHESSI X-ray count rates in different energy bands between 3 and 50 keV energies are represented during the M2.0 event. The GOES flux profile in the 1--8 \AA\ channel, AIA 193, and 304 \AA\ light curves of the flaring region are also overplotted. Panels (b)--(g): evolution of the flare shown in AIA 304 \AA\ images. The X-ray contours in 6--12, 12--25, and 25--50 keV are overplotted on EUV images. The X-ray images are reconstructed by the CLEAN algorithm with integration time of 20 s. The contours drawn are at 50\%, 60\%, 70\%, 80\%, and 90\% of the peak flux in each image. The yellow arrows (except in panel (d)) indicate the plasma eruptions originating from the core region.}
\label{fig:AIA_M2.0}
\end{figure}

\begin{figure}
\centering
\includegraphics[scale=1.2]{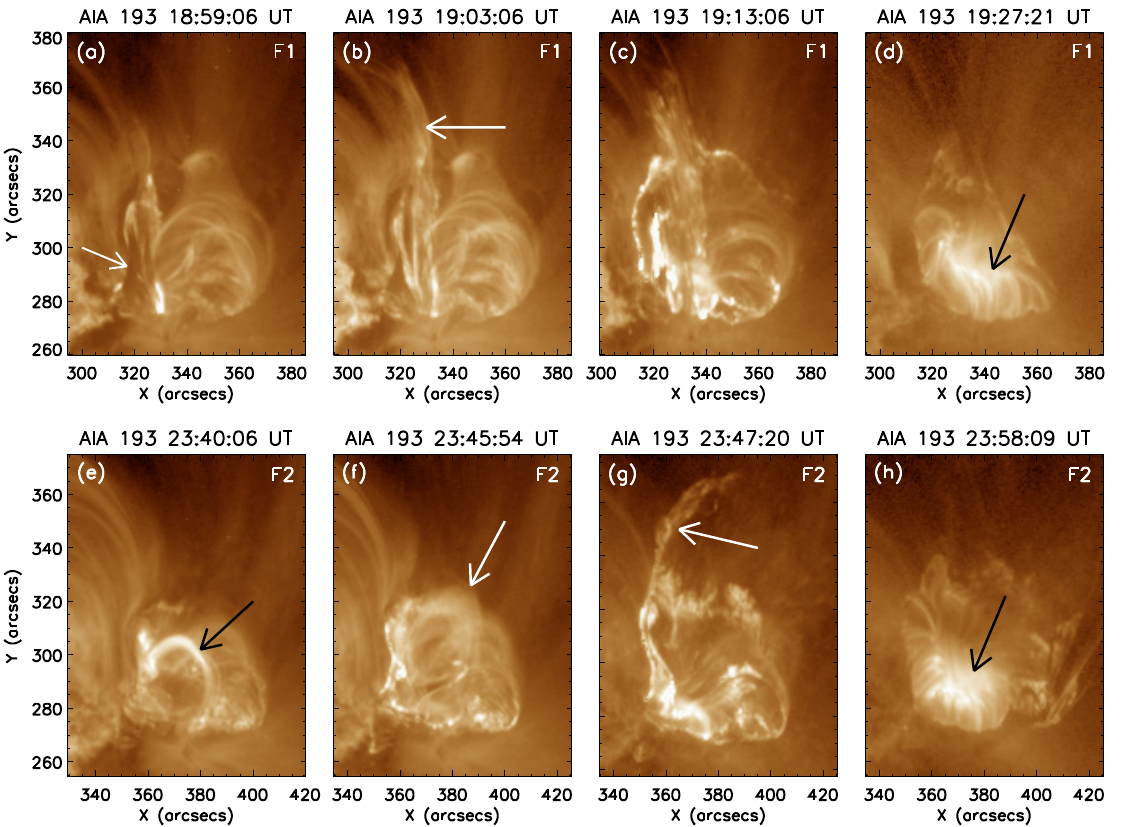}\caption{Panels (a)--(d): evolution of the M2.0 flare (F1) in the AIA 193 \AA\ images. In panel (a), we indicate an activated filament by an arrow with clear brightening at its base. Panel (b) shows the subsequent jet-like eruption of the filament. Panel (c) denotes the extended brightening within the flaring region, which marks the start of the impulsive phase of the flare. In panel (d), we indicate the compact post-flare loop system by an arrow. Panels (e)--(h): evolution of the M2.6 flare (F2) in the AIA 193 \AA\ images. In panel (e), we mark a bright loop system in the core region observed prior to the flare. This loop system erupts subsequently in a coherent manner, which we indicate by an arrow in panel (f). Thereafter, the erupting structure evolves non-coherently. We mark the bright eastern part of the erupting structure by an arrow in panel (g). We indicate the compact, bright post-flare loop arcades in panel (h) by an arrow.}
\label{fig:193_fig1}
\end{figure}

\begin{figure}[htp!]
\centering
\includegraphics[scale=0.7]{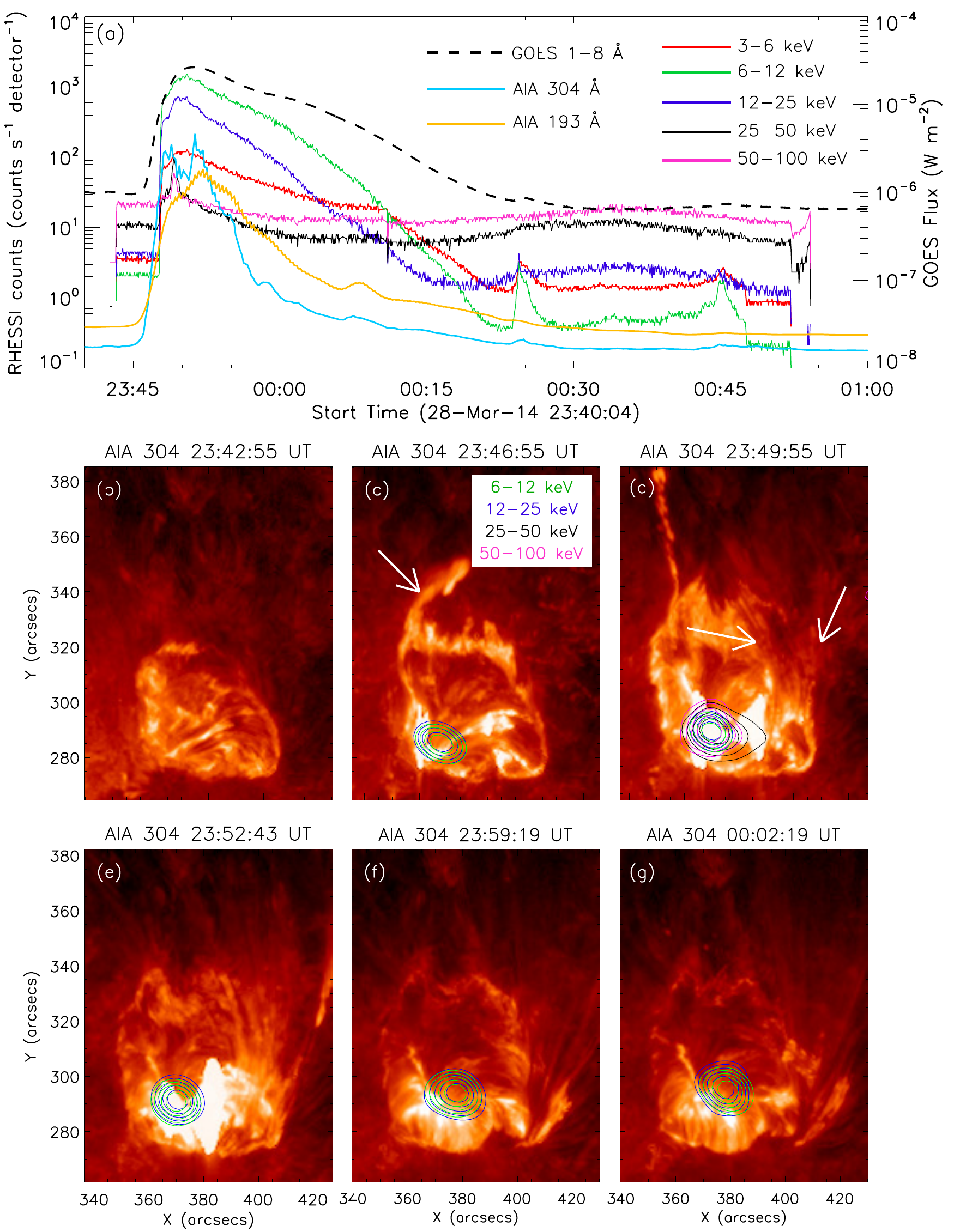}
\caption{Panel (a): RHESSI X-ray count rates in different energy bands between 3 and 100 keV are illustrated during the M2.6 event. The GOES flux profile in the 1--8 \AA\ channel, AIA 193, and 304 \AA\ light curves of the flaring region are also overplotted for comparison with the X-ray light curves. Panel (b): the preflare configuration of the flaring region observed in AIA 304 \AA, devoid of significant X-ray emissions. Panels (c)--(g): evolution of the flare in AIA 304 \AA\ observations along with the RHESSI X-ray sources overplotted on the EUV images. In panels (c)--(d), we indicate the plasma structures to be erupted from the eastern and western parts of the flaring region, respectively. The X-ray contours in 6--12, 12--25, 25--50, and 50--100 keV are overplotted on the EUV images. The X-ray images are reconstructed by the CLEAN algorithm with integration time of 20 s. The contours drawn are at 50\%, 60\%, 70\%, 80\%, and 90\% of the peak flux in each image.}
\label{fig:AIA_M2.6}
\end{figure}

\begin{figure}[htp!]
\centering
\includegraphics[scale=0.7]{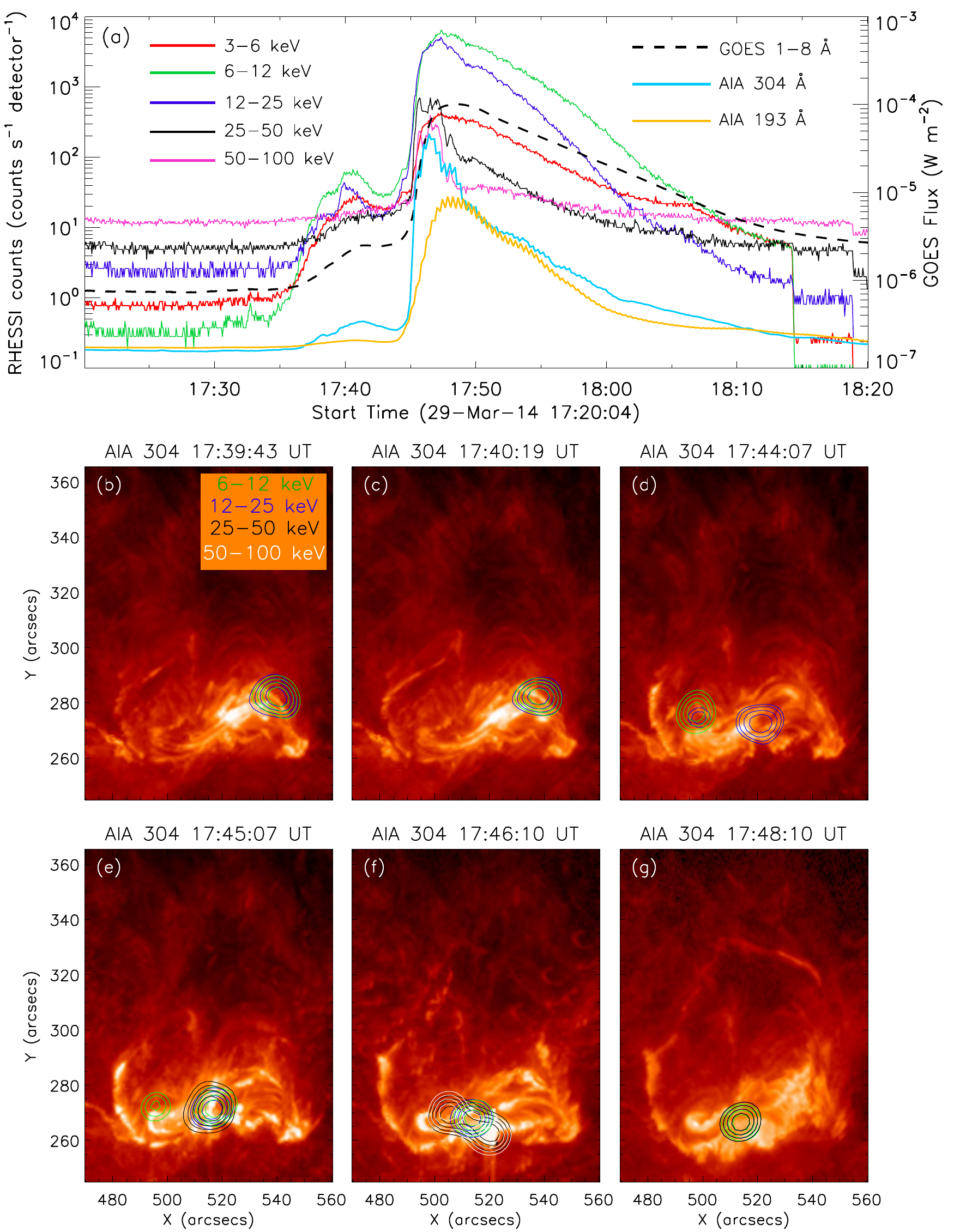}
\caption{Panel (a): RHESSI X-ray count rates in different energy bands between 3 and 100 keV are represented during the X1.0 event.  GOES flux profile in the 1--8 \AA\ chanel, AIA 193, and 304 \AA\ light curves of the flaring region are overplotted. Panels (b)--(g): AIA 304 \AA\ observations showing the evolution of the flare. X-ray contours in 6--12, 12--25, 25--50, and 50--100 keV are overplotted on the EUV images. The X-ray images are reconstructed by the CLEAN algorithm with integration time of 20 s. The contours drawn are at 60\%, 70\%, 80\%, and 90\% of the peak flux in each image.}
\label{fig:AIA_X1.0}
\end{figure}

\begin{figure}[htp!]
\centering
\includegraphics[scale=1.2]{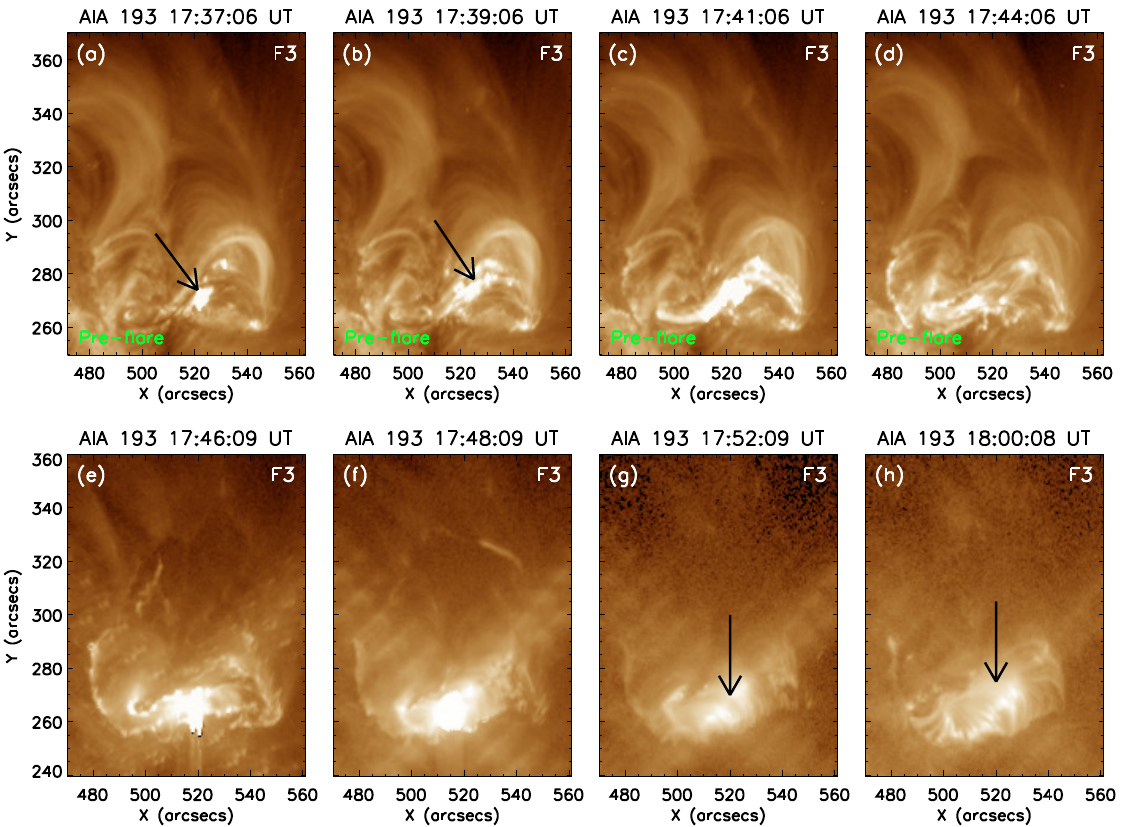}\caption{Evolution of the X1.0 flare (F3) in the AIA 193 \AA\ images. Panels (a)--(d) denote the morphological changes during the preflare phase of the flare, while the panels (e)--(h) correspond to the main phase of the flare. We note a preflare activity in the form of intense brightening from the western part of the core, which we indicate by arrows in panels (a)--(b). Subsequently, this preflare intensity enhancement spreads over a large area within the core region (panels (c)--(d)). During the main phase of the flare, enhanced intensity is observed from an extended part of the core region (panels (e)--(f)). Afterwards, the dense post-flare loop arcades are observed to form, which gradually get elongated over a broad area of the core, indicated by arrows in panels (g)--(h).}
\label{fig:193_fig2}
\end{figure}

\begin{figure}[htp!]
\includegraphics[width=\textwidth]{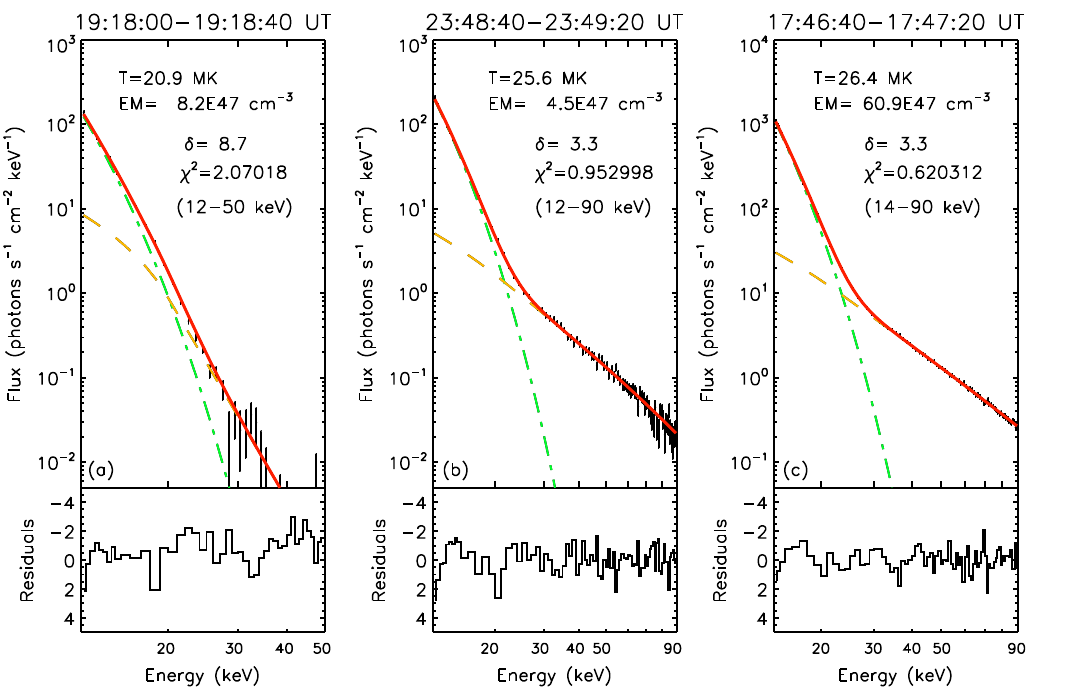}
\caption{Results of RHESSI X-ray spectral fit, along with their residuals, are shown for all the events under analysis. We used an isothermal model (shown by green dash-dotted line) for thermal fit and thick-target Bremsstrahlung model (shown by yellow dashed line) for non-thermal fit of the observed spectra. The red solid line indicates the sum of the two components. Each spectra was accumulated with an integration time of 40 s using the front segments of the detectors 1--9 (except 2 and 7). The energy ranges selected for the spectral fit are annotated in the respective panels. In panels (a), (b), and (c), we show the spectra at the peak of the HXR emissions during the M2.0, M2.6, and X1.0 flares, respectively.}
\label{fig:spectra}
\end{figure}

\begin{figure}[htp!]
\centering
\includegraphics[scale=1.2]{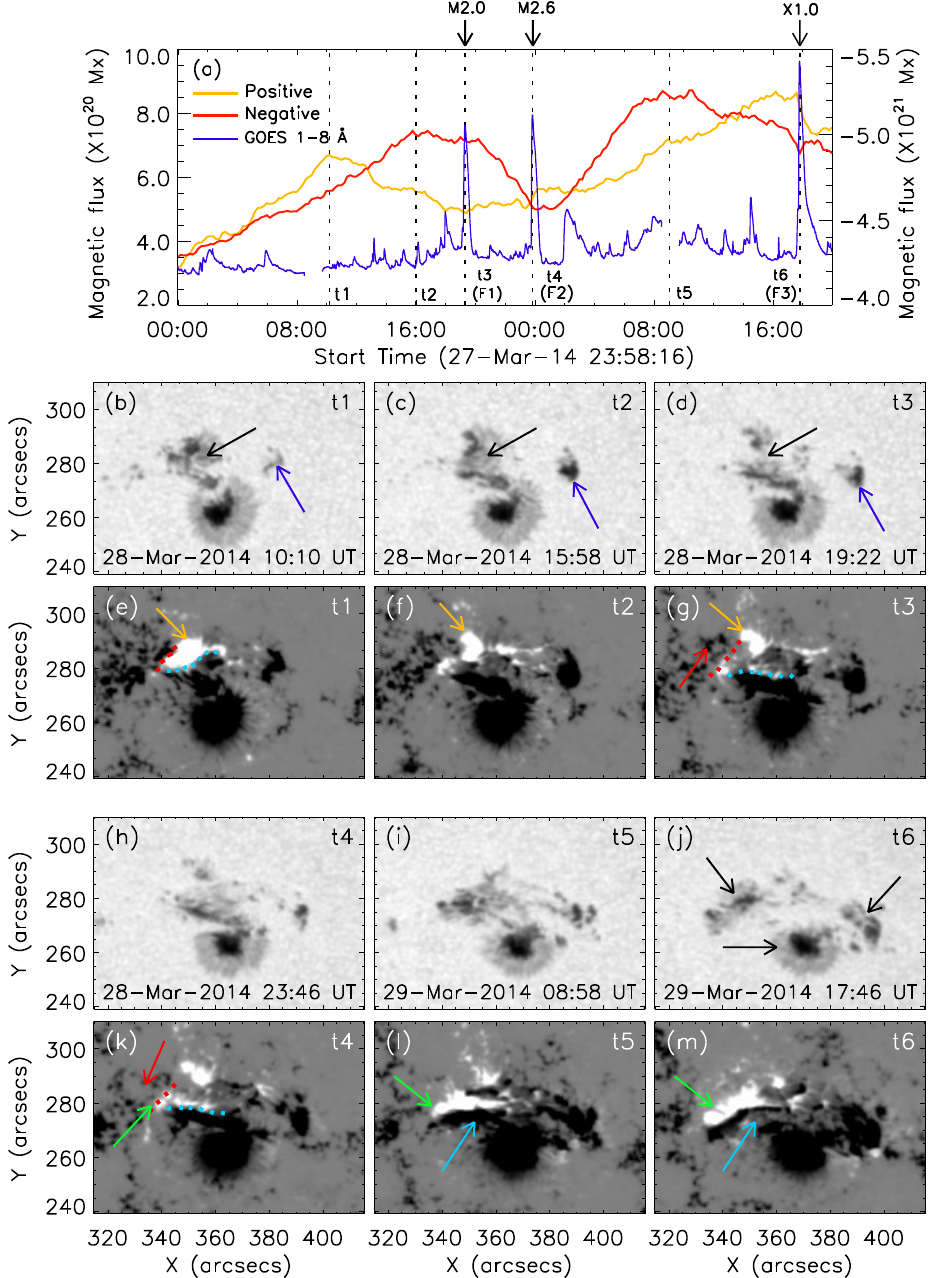}
\caption{Panel (a): the temporal evolution of magnetic flux obtained from the flaring region along with the GOES SXR light curve in 1--8 \AA\ channel. Panels (b)--(d): intensity images of the flaring region at three instances t1, t2, and t3 as marked in panel (a). The black arrows denote small-scale changes in the northern sunspot group, while the blue arrows indicate growth of a compact sunspot group. Panels (e)--(g): LOS magnetogram images cotemporal with continuum observations (shown in the previous row) at times t1, t2, and t3. The yellow and red arrows are used to indicate the changes in positive and negative fluxes, respectively. Panels (h)--(j): intensity images of the flaring region at t4, t5, and t6 as marked in panel (a). The arrows in panel (j) indicate three distinct sunspot groups. Panels (k)--(m): LOS magnetogram images for t4, t5, and t6 cotemporal with corresponding continuum observations (shown in the previous row). The green and sky blue arrows are used to indicate the changes in positive and negative fluxes, respectively. The red and sky-blue dotted lines in panels (e), (g), and (k) denote the PILs in the eastern and western parts of the flaring region, respectively.}
\label{fig:HMI}
\end{figure}

\begin{figure}[htp!]
\hspace{-0.3cm}
\includegraphics[scale=0.17]{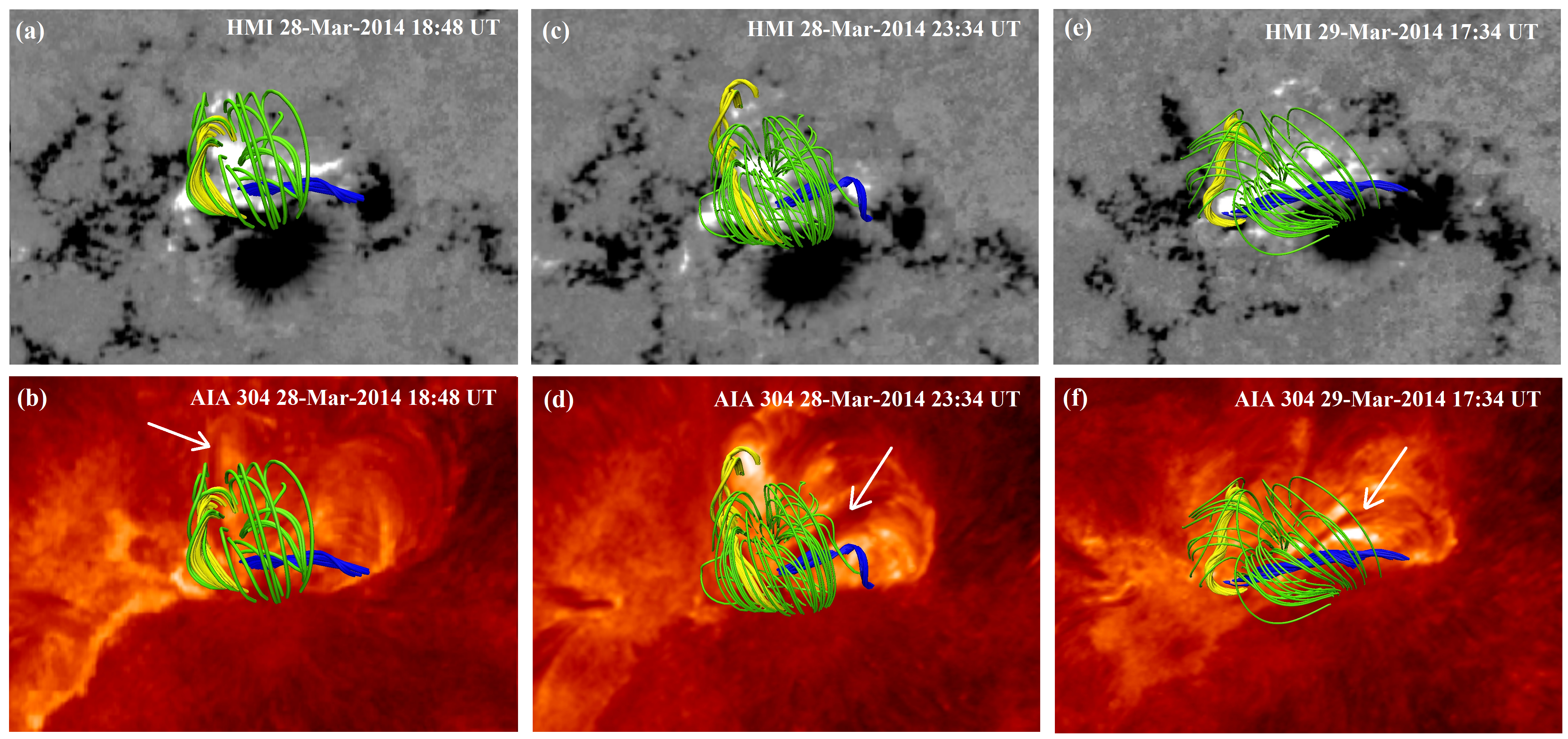}
\caption{The preflare coronal magnetic structures obtained through NLFFF modeling, presented over the photospheric radial magntic fields (top panels) and the AIA 304 \AA\ observations (bottom panels) in and around the flaring region for all the events. The panels (a)--(b), (c)--(d), and (e)--(f) denote the epochs corresponding to F1, F2, and F3, respectively. We note the existence of two flux ropes, shown by yellow and blue field lines, to reside over the eastern and western PILs of the flaring region (top panels; see Figure \ref{fig:HMI}), respectively. The flux ropes are enveloped by the low-lying bipolar green field lines. In the bottom panels, we mark the filaments by arrows, observed through EUV imaging.}
\label{fig:FR_304}
\end{figure}

\subsection{Temporal and spatial aspects}

The temporal and spatial evolution in EUV 304 \AA\ observations of the M2.0 (F1), M2.6 (F2), and X1.0 (F3) flares are presented in Figures \ref{fig:AIA_M2.0}, \ref{fig:AIA_M2.6}, and \ref{fig:AIA_X1.0}, respectively. The panel (a) in these figures presents light curves of the flares, while panels (b)--(g) show their spatial evolution. The temporal evolution of the flares has been studied with GOES 1--8 \AA, AIA 304 \AA, and RHESSI X-ray light curves. We have reconstructed RHESSI X-ray light curves in various energy bands, viz., 3--6, 6--12, 12--25, 25--50, and 50--100 keV. For F1, we do not show the 50--100 keV light curve, because of the lack of significant X-ray flux above 50 keV energy. To explore the spatial structures in the flaring region and their evolution at the upper chromospheric level, we plot a few representative AIA 304 \AA\ images overplotted by the RHESSI X-ray sources at various energy bands. We note that X-ray emission sources at 6--12 keV are exactly cospatial with the lower energy sources at 3--6 keV, hence not shown in these figures. To understand the evolution of the spatial structures at the flaring corona, we show the EUV 193 \AA\ images In Figures \ref{fig:193_fig1}(a)--(d), (e)--(h), and \ref{fig:193_fig2} for F1, F2, and F3, respectively.

\subsubsection{M2.0 flare}

Before the beginning of the rise phase of F1, we observe a preflare enhancement in the X-ray light curves at $\approx$19:08 UT. We note that this preflare hump is absent in the AIA 304 \AA\ light curve, which represents emission from the flaring region exclusively. This observation suggests that this preflare emission is not associated with the flaring event under analysis. We observe plasma eruption from the eastern part of the flaring region in the form of collimated stream (indicated by arrows in Figures \ref{fig:AIA_M2.0}(b)--(c)) at the outset of F1. At the base of the collimated structure, we note X-ray emission upto 25 keV, shown by contours of different energy bands. During the peak of the flare ($\approx$19:18 UT), the hard X-ray (HXR) source of 25--50 keV appeared at the flaring core (shown by black contours in Figure \ref{fig:AIA_M2.0}(d)). After the flare peak, we observe eruption of cool (i.e., dark) plasma from the western part of the core region (shown by arrows in Figures \ref{fig:AIA_M2.0}(e)--(g)). During this interval, the X-ray sources upto 25 keV energies are observed as a single source, suggesting X-ray production from a compact and dense system of coronal loops. The X-ray sources in the decay phase (Figures \ref{fig:AIA_M2.0}(f)--(g)) further confirm this scenario as the X-ray emission is observed to originate above the closely-packed post-flare loop system.

The evolutionary stages of F1 in the EUV 193 \AA\ images are shown in Figures \ref{fig:193_fig1}(a)--(d). Prior to the flare, we detect an activated filament (indicated by arrow in Figure \ref{fig:193_fig1}(a)) with clear signature of activity in the form of brightening at its base. Subsequently, the filament erupts in a jet-like manner (marked by arrow in Figure \ref{fig:193_fig1}(b)) with morphological similarity with the collimated stream observed in EUV 304 \AA\ images (see Figures \ref{fig:AIA_M2.0}(b)--(c)). The start of the impulsive rise phase of the flare can be discerned in the form of extended brightening over the flaring region (Figure \ref{fig:193_fig1}(c); see also Figure \ref{fig:AIA_M2.0}(a)). Afterwards, we observe the formation of compact post-flare loop arcades in the core region (shown by arrow in Figure \ref{fig:193_fig1}(d)).

\subsubsection{M2.6 flare}

The evolutionary stages of F2 are shown in Figure \ref{fig:AIA_M2.6} through the EUV 304 \AA\ image sequences. Like F1, in this case also, we observe a single X-ray source throughout the flare evolution. Furthermore, the X-ray sources from lower to higher energies (e.g., 6--12, 12--25, 25--50, and 50--100 keV) are observed to be cospatial. During the peak of the flare ($\approx$23:50 UT), the X-ray sources in the energy band of 50--100 keV is observed to appear in the core region (Figure \ref{fig:AIA_M2.6}(d)). Thereafter, the 50--100 keV source disappeared, while the X-ray emission in the lower energy bands persisted (Figures \ref{fig:AIA_M2.6}(e)--(g)). We note a double peak structure in the AIA 304 \AA\ light curve during the peak time of the flare (see Figure \ref{fig:AIA_M2.6}(a)). This double peak structure suggests two successive episodes of intense brightening that accompanies eruptions from the eastern and western parts of the flaring region, respectively (indicated by arrows in Figures \ref{fig:AIA_M2.6}(c)--(d)), which ultimately resulted into the flaring intensity of M2.6 class. Notably, the eruptions in this case are not jet-like ejections as in the case of F1.

In Figures \ref{fig:193_fig1}(e)--(h), we show the EUV 193 \AA\ images presenting the evolutionary stages of F2. Prior to the flare, we observe a bright, activated loop system in the core region (indicated by arrow in Figure \ref{fig:193_fig1}(e)). This loop structure erupts subsequently (marked by arrow in Figure \ref{fig:193_fig1}(f)) in a nearly coherent manner, which also marks the beginning of the impulsive phase of the flare (see Figure \ref{fig:AIA_M2.6}(a)). Thereafter, the erupting loop system loses its coherency and we mark its bright eastern part by an arrow in Figure \ref{fig:193_fig1}(g). Later on, the dense and compact post-flare loop arcades are observed to form in the core region (indicated by arrow in Figure \ref{fig:193_fig1}(h)).

\subsubsection{X1.0 flare}

The evolution of the X1.0 flare (F3) is shown by a few representative AIA 304 \AA\ images in Figures \ref{fig:AIA_X1.0}(b)--(g). Prior to the impulsive phase of the event, starting at $\approx$17:44 UT, we observe clear signature of preflare activity persisting for $\approx$8 min ($\approx$17:36--17:44 UT). This preflare phase is discernible in all the X-ray and EUV light curves (Figure \ref{fig:AIA_X1.0}(a)). This preflare activity is observed in the AIA images as enhanced brightening from the western part of the core region, which also emits X-ray sources upto 25 keV (Figures \ref{fig:AIA_X1.0}(b)--(c)). Thereafter, the X-ray sources evolved at two separate locations and we observe emission upto 25 keV from both the eastern and western parts of the flaring core region (Figure \ref{fig:AIA_X1.0}(d)). The X-ray emission in the 6--12 keV energy range persisted in the eastern part of the core, whereas, the strong emission in the energy range of 25--50 keV is observed to appear in the western part of the flaring core (Figure \ref{fig:AIA_X1.0}(e)). During the peak of HXR light curves, we observe clear X-ray sources in the 50--100 keV energy band. Importantly, this high energy source presents elongated structure with two distinct centroids (Figure \ref{fig:AIA_X1.0}(f)). Subsequently, the X-ray emission upto 50 keV persisted, observed as a single source structure in multiple energy bands (Figure \ref{fig:AIA_X1.0}(g)).

The evolution of F3 in the EUV 193 \AA\ images is presented in Figure \ref{fig:193_fig2}. Similar to the EUV 304 \AA\ observations, in these images also, we observe clear signature of enhanced brightening from the western part of the core revealing the preflare activity (indicated by arrows in Figures \ref{fig:193_fig2}(a)--(b)). Subsequently, this preflare enhancement spreads over the whole core region (Figures \ref{fig:193_fig2}(c)--(d)). During the main phase of the flare, intense and widespread flare emissions are observed from the core (Figures \ref{fig:193_fig2}(e)--(f)). Thereafter, the post-flare loop arcades are observed to form (shown by arrow in Figure \ref{fig:193_fig2}(g)), which gradually gets denser, brighter, and are observed to extend over a large area of the core (marked by arrow in Figure \ref{fig:193_fig2}(h)).


\subsection{RHESSI X-ray spectroscopy}
\label{sec:spectra}

To quantify the thermal and non-thermal components of X-ray emission during the three flares, we conduct X-ray spectroscopic analysis with RHESSI observations. We generate RHESSI spectra with an energy binning of 1/3 keV from 6 to 15 keV, 1 keV from 15 to 100 keV, and 5 keV from 100 keV onward. We use the front segments of the detectors 1--9 (except detectors 2 and 7, which have lower energy resolution and high threshold energy, respectively). The spectra are deconvolved with the full detector response matrix (i.e., off-diagonal elements are included; \citealp{Smith2002}). For thermal fitting, we use an isothermal model constructed using line spectrum. The non-thermal spectra are fitted using the thick-target bremsstrahlung model \citep{Holman2003}. We derive the temperature (T) and emission measure (EM) of the hot flaring plasma from the thermal fit and the electron spectral index ($\delta$) from the non-thermal component.

The results obtained from the spectral fit of X-ray emission from the flaring region are presented in Figures \ref{fig:spectra}(a), (b), and (c) for F1, F2, and F3, respectively. For F1, the GOES flare peak ($\approx$19:18 UT on 2014 March 28) coincides with the HXR (25--50 keV) peak. A high value of electron spectral index (i.e., $\delta$=8.7) indicates mild non-thermal component of flaring X-ray emission. During the peak of F2, the electron spectral index decreases to 3.3, indicating a much harder non-thermal spectrum compared to F1. From the thermal spectral fit, we obtain temperature of the flaring region as $\approx$25.6 MK, which is higher than the temperature ($\approx$20.9 MK) during the peak of F1. During F3, the hardness of the spectrum remains almost the same as that of F2. However, the emission measure ($\approx$61$\times$10$^{47}$ cm$^{-3}$) increases by an order of magnitude during F3 compared to the previous two events (see Figures \ref{fig:spectra}(a)--(c)). This indicates a significant enhancement in the electron density of hot (T $\approx$26 MK) plasma within the flaring volume.

\section{STRUCTURE AND EVOLUTION OF MAGNETIC FIELDS}
\label{sec:magnetic_evolution}


\subsection{Photospheric magnetic fields}
\label{sec:HMI_evolution}

We analyze the structural and temporal evolution of photospheric magnetic field of the flaring region in Figure \ref{fig:HMI}. To examine the magnetic flux changes quantitatively, we plot the spatial variations of positive and negative magnetic flux of the region of interest (shown in Figures \ref{fig:HMI}(b)--(m)). In Figure \ref{fig:HMI}(a), we provide the time profiles of the integrated magnetic fluxes through the selected area (see Figures \ref{fig:HMI}(b)--(m)) together with the GOES 1--8 \AA\ soft X-ray (SXR) light curve. The time profiles of magnetic fluxes start from 00:00 UT on 2014 March 28 to 20:00 UT on 2014 March 29 ($\approx$44 hr) covering all the flare events under analysis}. Also, the chosen interval includes a time span of $\approx$19 hr before F1 to examine the build-up of preflare photospheric flux in detail. 

In Figure \ref{fig:HMI}(a), we select six different epochs (t1, t2, t3, t4, t5, and t6) to explore the spatial changes in the photospheric magnetic field distribution. Among these epochs t3, t4, and t6 are selected at the peak time of the flares under analysis. The continuum and LOS magnetogram images during the epochs (t1--t6) are presented in Figures \ref{fig:HMI}(b)--(m). We observe substantial structural changes in the photospheric magnetic field of the flaring region over the selected interval of $\approx$44 hr (see Figures \ref{fig:HMI}(b)--(m)).

In Figures \ref{fig:HMI}(b)--(g), we show the evolution of flaring region with cotemporal continuum and magnetogram observations for three epochs t1, t2, and t3, which present magnetic field changes prior to F1. The inspection of these images reveals an increase followed by decrease of sunspot area in the northern sunspot group (shown by black arrows in Figures \ref{fig:HMI}(b)--(d)). We also observe growth of compact sunspot groups on the western side of the main sunspot group (indicated by dark-blue arrows in Figures \ref{fig:HMI}(b)--(d)). Figures \ref{fig:HMI}(e)--(g) show the cotemporal LOS magnetogram observations corresponding to continuum images in Figures \ref{fig:HMI}(b)--(d). In Figures \ref{fig:HMI}(e) and (g), we focus on the eastern and western PILs marked by red and sky-blue dotted lines, respectively. The yellow arrows indicate a gradual decrease of positive flux near the eastern PIL (Figures \ref{fig:HMI}(e)--(g)), whereas the red arrows indicate subsequent decrease of negative flux (see Figures \ref{fig:HMI}(g) and (k)). We further note that the orientation of the western PIL has changed from t1 to t3 (see the sky-blue dotted lines in Figures \ref{fig:HMI}(e) and (g)). 

In Figures \ref{fig:HMI}(h)--(m), we present the continuum and LOS magnetogram observations showing the evolution of the flaring region during the epochs t4, t5, and t6, which are used to explore the changes in the photospheric magnetic field structures associated with F2 and F3. Notably, after F2, the photospheric configuration of sunspot groups and associated magnetic fields at the northern region exhibit striking changes (see Figures \ref{fig:HMI}(h)--(j)). We observe that, during the interval between F2 and F3, the northern sunspot group with relatively compact configuration undergoes rapid expansion, which resulted in its fragmentation into three distinct parts (indicated by arrows in Figure \ref{fig:HMI}(j)). The magnetogram images cotemporal with the continuum observations are shown in Figures \ref{fig:HMI}(k)--(m). The eastern and western PILs are indicated by red and sky-blue dotted lines, respectively in Figure \ref{fig:HMI}(k). We observe clear features of flux cancellation and emergence. The sky-blue arrows indicate a substantial cancellation of negative flux near the western PIL, whereas the green arrows indicate a gradual increase of positive flux near the same PIL. These observations showing flux emergence and cancellation are in agreement with the qualitative estimation shown in Figure \ref{fig:HMI}(a).

\subsection{Magnetic configuration of flaring corona}
\label{sec:mag_corona}

In Figure \ref{fig:FR_304}, we represent the preflare coronal magnetic structures in and around the flaring region obtained through NLFFF extrapolation. The first, second, and third columns denote the epochs corresponding to F1, F2, and F3, respectively. In the top panels, we show the extrapolated coronal field lines taking the HMI SHARP CEA radial magnetic fields as the background, whereas, in the bottom panels, we show the AIA 304 \AA\ images in the background using the visualization software VAPOR \citep{Li2019}. In all the cases, we find existence of two flux ropes lying over the compact eastern and western PILs of the flaring region (see Figure \ref{fig:HMI}), shown by yellow and blue field lines, respectively. The flux ropes are encompassed by the low-lying bipolar field lines (shown in green). The sequential eruptions of the flux ropes and neighboring core field gave rise to the eruptive flares under analysis. We observe filaments in the AIA 304 \AA\ images, which are indicated by arrows in the bottom panels. The detailed investigation on the modeling of coronal magnetic fields during the three events is in progress and will be presented in a subsequent study.

\begin{figure}[htp!]
\centering
\includegraphics[scale=1.3]{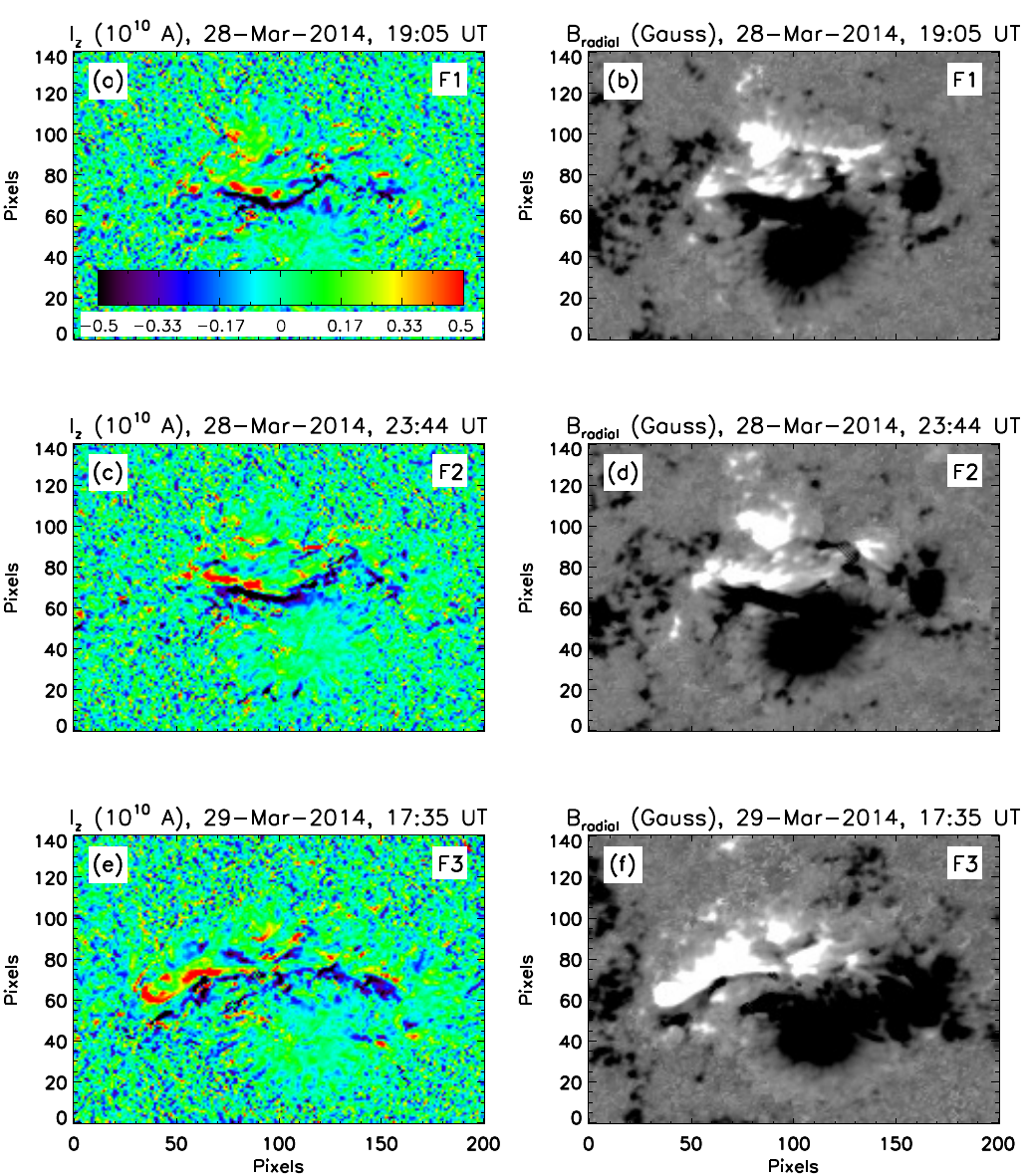}
\caption{The comparison of photospheric longitudinal current (I$_{z}$; first column) and the radial component of the magntic field (B$_{radial}$; second column) before the onset of the flares. The panels (a)--(b), (c)--(d), and (e)--(f) correspond to F1, F2, and F3, respectively. In panel (a), we denote the colorbar corresponding to the distribution of I$_{z}$ in panels (a), (c), and (e). We saturate the I$_{z}$ and B$_{radial}$ values to $\pm$0.5$\times$10$^{10}$A and $\pm$500G, respectively in all the panels. We note that the current distribution is strong near the western PIL of the flaring region (see Figure \ref{fig:HMI}) and it gets elongated along the PIL before F3. The morphological changes in the longitudinal current distribution are similar to the structural changes in the distribution of the radial magnetic field component (see panels (e)--(f)).}
\label{fig:current}
\end{figure}

\begin{figure}[htp!]
\centering
\includegraphics[scale=0.15]{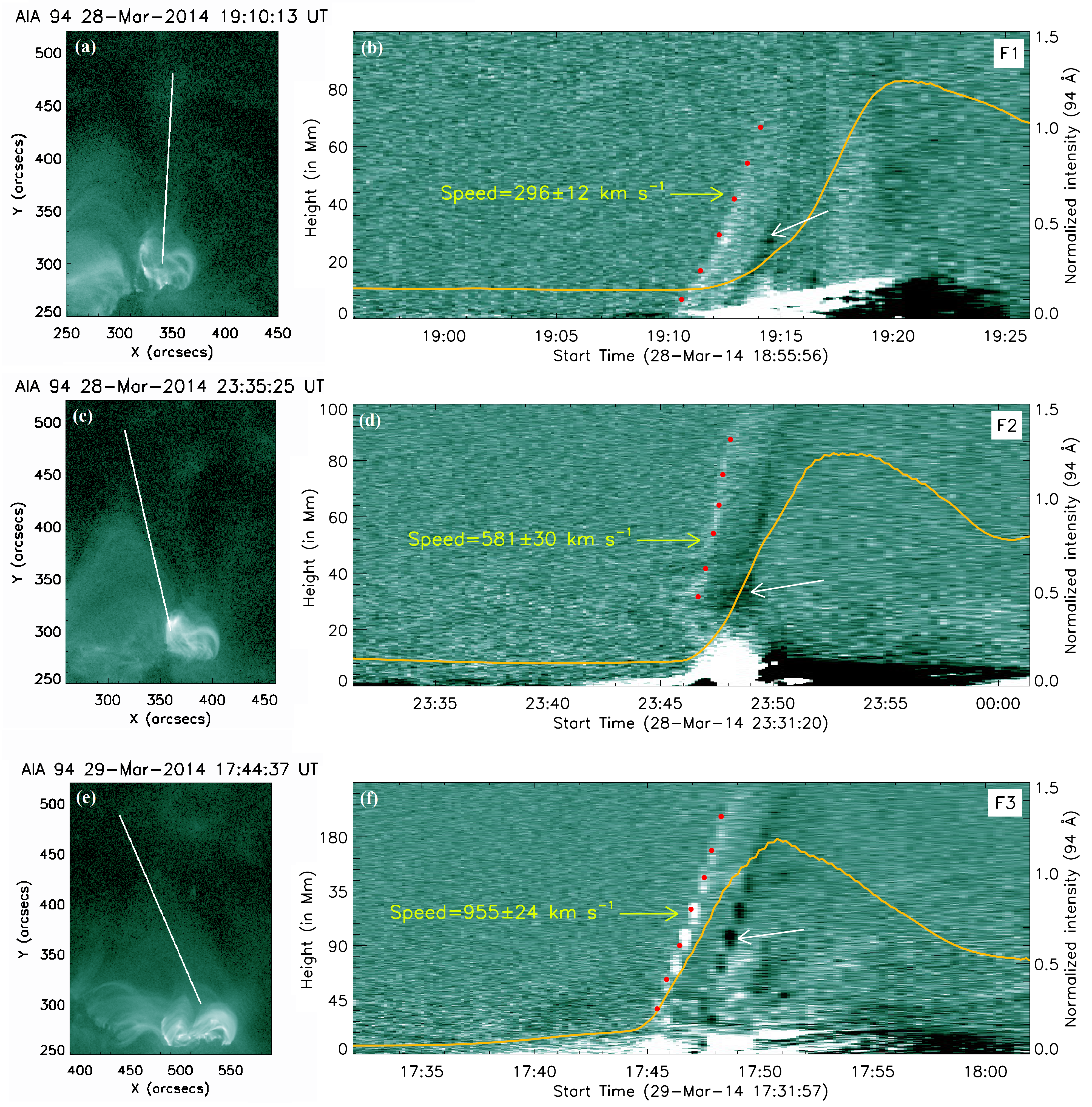}
\caption{The evolutionary stages of the eruption of hot plasma structures from the core flaring region. The panels (a)--(b), (c)--(d), and (e)--(f) correspond to F1, F2, and F3, respectively. The left column shows the direction of the slits over AIA 94 \AA\ direct images, along which we calculated the height-time profiles of the eruptions. In the right column, we show the time-slice diagrams obtained from the AIA 94 \AA\ running difference images by tracking the intensity variation along the slits. The erupting hot plasma structures are indicated by red dots in panels (b), (d), and (f). The speeds of the erupting structures are annotated in these panels with the corresponding uncertainty in the measurements. We note that the speeds are gradually increasing from F1 to F3. Following the hot plasma eruption, we observe the eruption of a dark (i.e, cool) structure, which we mark by white arrows in these panels. The flare intensity profiles in the AIA 94 \AA\ channel are also overplotted by orange curves.}
\label{fig:time_slice}
\end{figure}



\section{Morphology and evolution of photospheric longitudinal currents}
\label{sec:current}

We show the morphological changes associated with the photospheric longitudinal current in response to the magnetic field changes in Figure \ref{fig:current}. On the photosphere, the vertical component of the electric current density (i.e., J$_{z}$) can be obtained from the horizontal magnetic field components (i.e., B$_{x}$ and B$_{y}$) using the Ampere's law \citep{Kontogiannis2017,Fleishman2018,Fursyak2020}:

\begin{equation}
J_{z}=\frac{1}{\mu_{0}}\Big(\frac{dB_{y}}{dB_{x}}-\frac{dB_{x}}{dB_{y}}\Big).
\end{equation}

The magnetic field components (B$_{z}$, -B$_{y}$, B$_{x}$) in Heliographic Cartesian Coordinate can be approximately obtained from the corresponding field components (B$_{r}$, B$_{\theta}$, B$_{\phi}$) in the Heliocentric Spherical Coordinate \citep{Gary1990}. In order to calculate the longitudinal current (I$_{z}$) from longitudinal current density (J$_{z}$), we need to multiply J$_{z}$ with the area of one pixel, i.e., 13.14$\times$10$^{10}$ m$^{2}$. In Figure \ref{fig:current}, we present the distribution of current (i.e., I$_{z}$) along with the structure of the radial component of magnetic field (i.e., B$_{r}$) within the flaring region before the start of the flares. The top, middle, and bottom panels of Figure \ref{fig:current} correspond to F1, F2, and F3, respectively. For better visualization, we saturate the current values at $\pm$0.5$\times$10$^{10}$ A in all the panels. The color code for all the I$_{z}$-maps (Figures \ref{fig:current}(a), (c), and (e)) is shown by a colorbar in Figure \ref{fig:current}(a). We observe significant amount of current concentration along the western PIL (see the sky-blue dotted lines in Figures \ref{fig:HMI}(e), (g), and (k)) of the flaring region for all events. Notably, the negative current largely dominates the positive current in all the cases. Between F2 and F3, the western PIL undergoes elongation (Figure \ref{fig:current}(f); see also Figures \ref{fig:HMI}(k)--(m)). In a similar way, the region of strong photospheric currents that predominantly exists at the flaring core region, is observed to show an extended morphological structure prior to F3 (see Figures \ref{fig:current}(a), (c), and (e)).



\begin{figure}[t!]
\centering
\includegraphics[scale=0.9]{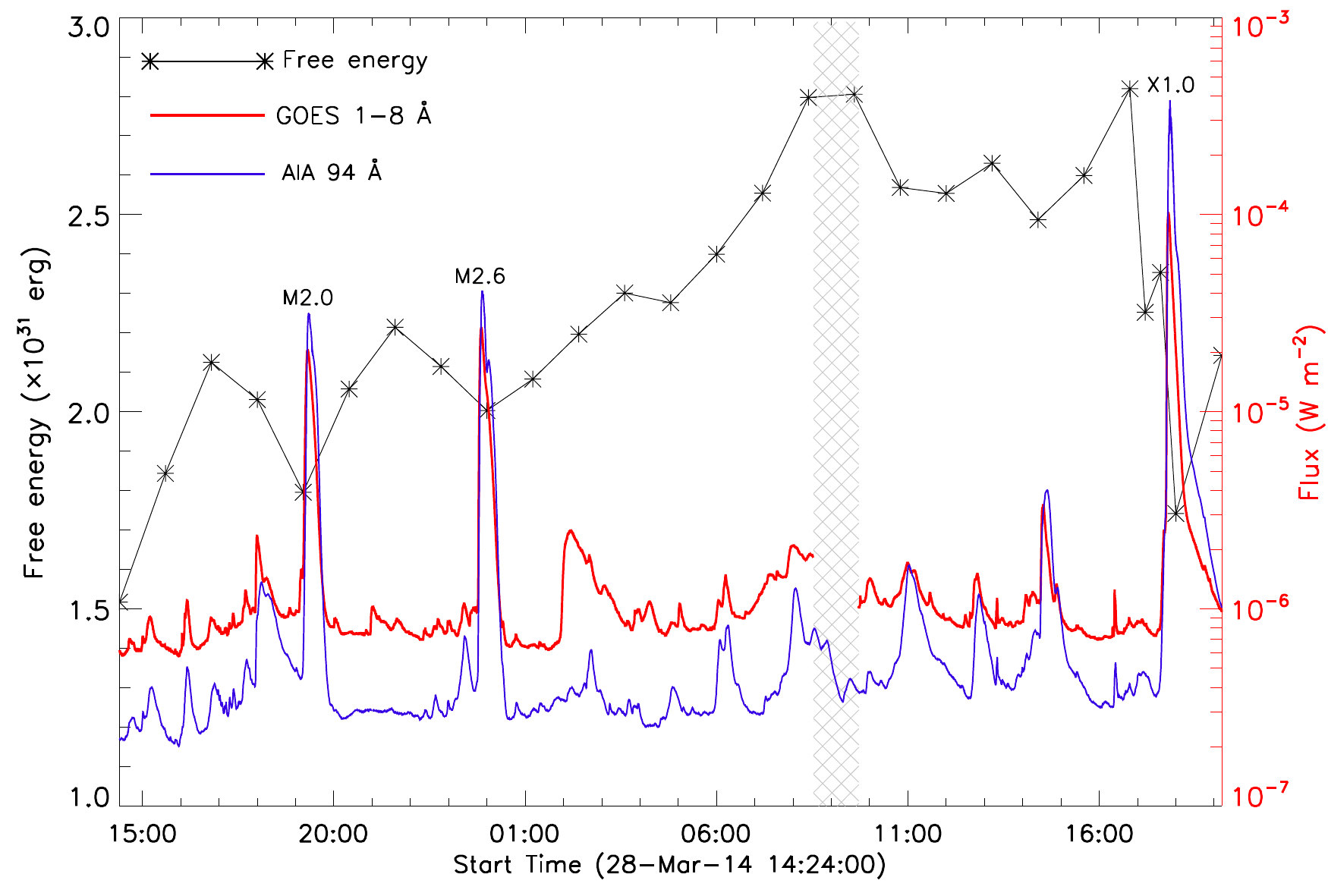}
\caption{The evolution of free magnetic energy during an interval of $\approx$30 hr encompassing all the flares under study. The free energy was calculated in the coronal volume encompassing the AR, taking the HMI SHARP CEA cutout as the photospheric boundary. We also plot full disk GOES light curve in the 1--8 \AA\ channel. The hatched region shows an interval during which GOES data were unavailable. To complement the data gap, we plot EUV 94 \AA\ intensity curve obtained from the AIA recorded from the active region.}
\label{fig:free_energy}
\end{figure}

\section{Onset of eruption}
\label{sec:time_slice}

For a quantitative understanding of the eruption from the flaring core, we present the time-slice diagrams in Figure \ref{fig:time_slice}. The Figures \ref{fig:time_slice}(a)--(b), (c)--(d), and (e)--(f) correspond to F1, F2, and F3, respectively. In the left column of this figure, we show the direction of the slits, along which we observed significant eruptive features. In the right column, we show the corresponding time-slice diagrams in AIA 94 \AA\ running difference images, constructed by tracking the eruptive signatures along the slits. We mark the erupting hot plasma structures by red dots in Figures \ref{fig:time_slice}(b), (d), and (f). We note that the speeds of the hot plasma ejections show an increasing trend from F1 to F3 (i.e., $\approx$296, 581, and 955 km s$^{-1}$). Following the hot plasma eruption, an eruption of dark (i.e., cool) material ensues, which we indicate by white arrows. To compare the eruption of the plasma structure with the temporal evolution of the flare, we overplot the AIA 94 \AA\ flare light curves in all the panels. We note that the rise of the 94 \AA\ intensity is near simultaneous with the onset of the hot plasma eruption from the flaring core. Subsequently, the eruptions from the source region resulted into CMEs. According to LASCO CME catalog\footnote{\url{https://cdaw.gsfc.nasa.gov/CME_list/UNIVERSAL/2014_03/univ2014_03.html}}, the linear speed of CMEs within LASCO field of view (FOV) associated with F1, F2, and F3 are 420, 503, and 528 km s$^{-1}$, respectively.

\section{Evolution of free magnetic energy} 
\label{sec:free_energy}

Our analysis is comprised of holomologous eruptive flares, which show gradully increasing SXR intensities (M2.0, M2.6, and X1.0). To investigate the scenario in the framework of storage and release process of free magnetic energy associated with the complex magnetic configuration of the AR, we calculate the temporal evolution of free energy during an interval of $\approx$30 hr, which is demonstrated in Figure \ref{fig:free_energy}. The free magnetic energy (E$_{F}$) is defined as the difference between the non-potential (E$_{NP}$) and potential (E$_{P}$) magnetic energies. i.e., 
\begin{equation}
          E_{F}=E_{NP}-E_{P},
\end{equation}

where E$_{NP}$ is calculated from the magnetic fields obtained from the NLFFF extrapolation. The different forms of energies can be calculated from the magnetic field information using the following relation

\begin{equation}
E=\int_V \frac{{B}^2}{8\pi} \,dV.
\end{equation}

We observe that there is prominent decrease of free magnetic energy due to the occurrence of the flaring events. We calculate the decrease to be 17\%, 9.5\%, and 38\% for the events M2.0, M2.6, and X1.0, respectively. There is a data gap in GOES light curve (in the 1--8 \AA\ channel) from 08:30 UT to 09:40 UT on 2014 March 29, which is indicated by a hatched region. To confirm if the GOES SXR flux enhancements are associated with the flaring activity in the AR 12017 of our interest, we have shown the AIA 94 \AA\ light curve deduced exclusively from the AR. It is observed, that in general, the EUV light curve matches well with the GOES light curve and the prominent SXR peaks represent activity in the AR.

\section{DISCUSSION}
\label{sec:discussion}







In this study, we explore the multi-wavelength evolution of three homologous flaring events of successively increasing intensities and associated energy release processes. The events occurred during 2014 March 28--29 in NOAA AR 12017 within an interval of $\approx$24 hr. The flares were triggered by the eruptions of flux ropes from the core of the AR. The importance of this study lies in the investigation of the intrinsic coupling of magnetic fields and associated processes from the photosphere to the corona that resulted into the repetitive build-up of compact magnetic flux ropes and their subsequent eruptions, observed in the form of homologous eruptive flares of successively increasing intensities. The important observational aspects of this study are summarized as follows:



\begin{enumerate}

\item According to the GOES Observations, the duration of the flares of our analysis are 22, 14, and 19 minutes for M2.0 (F1), M2.6 (F2), and X1.0 (F3) flares, respectively (see Table \ref{tab:summary}). A statistical analysis of almost 50,000 GOES SXR flares during the period 1976--2000 was presented in \citet{Veronig2002}. Their analysis revealed that the average values of the duration for M and X class flares are 24 and 30 minutes, respectively. In view of this, the duration of F1 is close to the value as revealed in the study of \citet{Veronig2002}, while F2 and F3 are of shorter duration. Notably, although F1 and F2 are of comparable intensity and homologous in nature, the duration of F2 is significantly smaller than that of F1. On the other hand, F3, despite being a large X-class flare, exhibits shorter duration compared to F1.

\item Inspection of the RHESSI X-ray images in multiple energy bands within the energy range of 3--100 keV during the evolution of F1 and F2 reveals a single X-ray source persisting throughout the flaring intervals. In both the events, the X-ray emissions are observed to come from the dense and closely-packed coronal loop system. The X-ray observations during the third event reveal emissions from different spatial locations within the flaring region (see Figure \ref{fig:AIA_X1.0}). During the onset of the impulsive phase of the third flare, we observe X-ray emissions from both the eastern and western parts of the flaring region (Figures \ref{fig:AIA_X1.0}(d)--(e)). Notably, conjugate X-ray sources with two distinct centroids in the 50--100 keV energy range are observed within the core region during the peak of the X1.0 flare (Figure \ref{fig:AIA_X1.0}(f)). The origin of these conjugate X-ray sources is likely to be caused by the deposition of energy by the energetic electrons at the footpoints of the post-reconnected loop system, as depicted in the standard flare model (see e.g., \citealp{Joshi2009}).

\item {We observe significant cancellation of magnetic fluxes (both positive and negative) within a bipolar flaring core region, near the eastern and western PILs (see Figure \ref{fig:HMI} and Section \ref{sec:HMI_evolution}). A detailed comparison of HMI magnetograms and EUV images (see Figures \ref{fig:AIA_M2.0}--\ref{fig:193_fig2} and Figure \ref{fig:HMI}) reveals that the source region of eruptions is spatially well correlated with the compact PILs (see Figure \ref{fig:FR_304}). Notably, we observe significant flux changes near the eastern PIL before F1 and F2, whereas, the change of flux is maximum near the western PIL before F3 (see Figure \ref{fig:HMI}). Thus, our observations imply a precise link between each flux rope eruption and magnetic flux cancellation at the photosphere. The observations of extended phases of flux cancellation together with prominent flux cancelling features near the PIL have important implications to understand the repetitive build-up of flux ropes and the triggering mechanism of the homologous eruptions. Contextually, from EUV imaging observations, we note that the eruptions initiated from the eastern PIL during F1 and F2, whereas, during F3, the eruption gets triggered from the western PIL. These observations show conformity with the tether-cutting model of solar eruption \citep{Moore1992,Moore2001}, where the build-up process of a flux rope along the PIL is governed by the flux cancellation that extends over a much longer interval compared to the flare time-scales. Pre-existence of magnetic flux rope(s) in ARs in relation to the eruptive flares and CMEs is well recognized (see e.g., \citealp{shibata1999}). Several contemporary observations have also confirmed the slow activation of a magnetic flux rope, prior to the flare's impulsive phase (e.g., \citealp{Joshi2016,Mitra2019,Sahu2020,Kharayat2021}). In our work, for each eruption, the NLFFF extrapolation results reveal the presence of two flux ropes, corresponding to the eastern and western PILs. The synthesis of EUV imaging, magnetogram observations, and NLFFF modeling suggests that the formed flux rope was likely destabilized by rapidly evolving, localized magnetic field structures near the PIL, thus indicating the role of small-scale tether-cutting reconnection in the triggering process. This scenario is further confirmed by the location of X-ray sources during the early phase, when the emission originates only at lower energies (below 25 keV; see Figures \ref{fig:AIA_M2.0}(b), \ref{fig:AIA_M2.6}(c), and \ref{fig:AIA_X1.0}(b)), as these sources are well correlated with the locations of prominent flux cancellation.}

\item {We further highlight the results of NLFFF extrapolation that reveal the presence of two magnetic flux ropes, formed over the eastern and western PILs prior to each eruptive flare. In this context, it is relevant to discuss and compare different aspects related to the MFR manifestations in this AR studied in detail by \citet{Yang2016} and \citet{Woods2018} in association with the flaring activity of March 28--29. Using a NLFFF model, \citet{Yang2016} showed a magnetic flux rope in the region for which twist number and decay index of the constraining field were calculated. Their analysis revealed that decay index lies below the critical value for torus instability to be operational. Therefore, the authors favoured the role of twist instability toward the CME onset although they could not find a common critical value of the twist number over which the flux rope tends to erupt. Nevertheless, their work points toward the fact that the twist number is a sensitive parameter in relation to flare occurrence. In a subsequent work, \citet{Woods2018} studied the most energetic flare (X1.0 event) from this AR which is the third event of our study. They found that the flux rope was actually composed of two flux ropes, of which only one erupted during the X1.0 flare. The detailed EUV and X-ray imaging observations presented in our work also reveal the destabilization of flux rope formed at the western PIL, in agreement with the work of \citet{Woods2018}. Because of the presence of the magnetic flux cancellation and brightening below the flux rope, \citet{Woods2018} concluded that the tether-cutting mechanism was responsible for the rising of the western flux rope to a torus unstable region prior to the flare. From the flare light curves along with corresponding EUV imaging observations, we clearly find preflare brightening for X1.0 flare (see Figures \ref{fig:AIA_X1.0}(a)--(d)), which further points toward the role of tether-cutting reconnection for the activation of flux rope to the torus unstable region. For a series of four eruptive events, similar result was obtained by \citet{Mitra2021}, where they proposed the combined role of ideal (torus) and resistive (tether-cutting) instability for the onset of CMEs.}

\item {The sequential eruptions of the flux ropes from the flaring core gave rise to corresponding CMEs. The inspection of the series of AIA 94 \AA\ running difference images reveals eruptions of hot coherent plasma structures (i.e., heated flux ropes) from the core region (Figure \ref{fig:time_slice}). We note that the speed of the flux rope eruptions in the source region, significantly increases from F1 to F3. The comparison of the speeds of eruptive flux ropes at the source region with the corresponding CMEs in LASCO FOV (see Section \ref{sec:time_slice}) reveals that the first flux rope undergoes acceleration (296 vs 420 km s$^{-1}$), the second one moves with approximately constant speed (581 vs 503 km s$^{-1}$), while the last eruption exhibits deceleration (955 vs 528 km s$^{-1}$)}.

\item {The build-up of electric current on the photosphere is directly associated with the emergence of current-carrying flux \citep{Tan2006,Torok2014}; their study is important to understand the build-up of non-potentiality in the active region corona \citep{Schrijver2005,Dalmasse2015}. Our study reveals that a strong current accumulation occurs near the western PIL of the flaring region (Figure \ref{fig:current}), where one of the two flux ropes lies (Figure \ref{fig:FR_304}). Furthermore, before F3, the magnetic flux near the western PIL is observed to undergo expansion showing an extended morphology (Figure \ref{fig:current}(f)). In response to this, the longitudinal current distribution gets elongated along the same PIL before F3 (Figure \ref{fig:current}(e)). In general, the role of photospheric currents have important consequences in triggering solar eruptive events. \citet{Mitra2020} studied the role of precursor flare activity in triggering a dual-peak M-class flare. Their study revealed the presence of strong, localized regions of photospheric currents of opposite polarities at the precursor location, making the region susceptible to small-scale magnetic reconnection}.

\item {The photospheric flux emergence and shearing motion introduce strong electric currents and inject energy into the active region corona. The coronal fields get reconfigured in this process and result into the accumulation of free magnetic energy in the coronal volume \citep{Regnier2012,Vekstein2016}. This stored free energy is regarded as a prime factor responsible for the explosive phenomena. The successively increasing intensities of the homologous flares of our analysis point toward a complex `storage and release' process of magnetic energies in the AR. 
For a quantitative understanding of the energy storage and release process, we study the evolution of free magnetic energy for a time span covering the three homologous events (Figure \ref{fig:free_energy}).
We find that the  maximum release of free magnetic energy (i.e., 38\%) is observed during the strongest event (i.e., F3/X1.0 flare). It is also remarkable to notice that the third event got a prolonged period for the storage of free magnetic energy (i.e., a period of $\approx$17 hr between F2 and F3), during which no major flare above class C occurred in the AR. Interestingly, this `storage phase' largely overlaps with a peristent phase of flux emergence (see Figure \ref{fig:HMI}(a); the interval between t4 and t5). In conclusion, our analysis reveals that the dominant variation in magnetic flux (both at large-scale involving the full flaring region as well as small-scales close to the compact PIL) and build-up of free magnetic energy in and around the flaring region is the root cause for the homologous eruptive flares of successively increasing intensities.}



\end{enumerate}

In summary, our paper provides a detailed investigation of multi-wavelength evolution of three homologous eruptive flares by combining HXR, EUV, white light, and magnetogram observations. We provide a quantitative estimation of the evolution of free magnetic energy in the corona associated with the AR, and explore its link with the ongoing photospheric and coronal processes. 
Thus, our study brings out the link between the photospheric developments resulting into the rapid build-up and subsequent eruption of coronal magnetic structures.

\acknowledgements

The authors would like to thank the SDO and RHESSI teams for their open data policy. SDO is NASA's mission under the Living With a Star (LWS) program. RHESSI was NASA's mission under the SMall EXplorer (SMEX) program. HMI data are courtesy of NASA/SDO and the HMI science team. A.P. would like to acknowledge the support by the Research Council of Norway through its Centres of Excellence scheme, project number 262622, as well as through the Synergy Grant No 810218 459 (ERC-2018-SyG) of the European Research Council. A.P. would also like to acknowledge partial support from NSF award AGS-2020703. The authors thank the anonymous referee for providing constructive comments and suggestions, which improved the overall quality of the article.

\bibliography{ms.bib}{}

\begin{thebibliography}{}
\expandafter\ifx\csname natexlab\endcsname\relax\def\natexlab#1{#1}\fi

\bibitem[{{Benz}(2017)}]{Benz2017}
{Benz}, A.~O. 2017, Living Reviews in Solar Physics, 14, 2

\bibitem[{{Chatterjee} \& {Fan}(2013)}]{Chatterjee2013}
{Chatterjee}, P., \& {Fan}, Y. 2013, \apjl, 778, L8

\bibitem[{{Cheung} {et~al.}(2019){Cheung}, {Rempel}, {Chintzoglou}, {Chen},
  {Testa}, {Mart{\'\i}nez-Sykora}, {Sainz Dalda}, {DeRosa}, {Malanushenko},
  {Hansteen}, {De Pontieu}, {Carlsson}, {Gudiksen}, \& {McIntosh}}]{Cheung2019}
{Cheung}, M.~C.~M., {Rempel}, M., {Chintzoglou}, G., {et~al.} 2019, Nature
  Astronomy, 3, 160

\bibitem[{{Dalmasse} {et~al.}(2015){Dalmasse}, {Aulanier}, {D{\'e}moulin},
  {Kliem}, {T{\"o}r{\"o}k}, \& {Pariat}}]{Dalmasse2015}
{Dalmasse}, K., {Aulanier}, G., {D{\'e}moulin}, P., {et~al.} 2015, \apj, 810,
  17

\bibitem[{{DeVore} \& {Antiochos}(2008)}]{Devore2008}
{DeVore}, C.~R., \& {Antiochos}, S.~K. 2008, \apj, 680, 740

\bibitem[{{Fleishman} \& {Pevtsov}(2018)}]{Fleishman2018}
{Fleishman}, G.~D., \& {Pevtsov}, A.~A. 2018, in Electric Currents in Geospace
  and Beyond, ed. A.~{Keiling}, O.~{Marghitu}, \& M.~{Wheatland}, Vol. 235,
  43--65

\bibitem[{{Fletcher} {et~al.}(2011){Fletcher}, {Dennis}, {Hudson}, {Krucker},
  {Phillips}, {Veronig}, {Battaglia}, {Bone}, {Caspi}, {Chen}, {Gallagher},
  {Grigis}, {Ji}, {Liu}, {Milligan}, \& {Temmer}}]{Fletcher2011}
{Fletcher}, L., {Dennis}, B.~R., {Hudson}, H.~S., {et~al.} 2011, \ssr, 159, 19

\bibitem[{{Fursyak} {et~al.}(2020){Fursyak}, {Kutsenko}, \&
  {Abramenko}}]{Fursyak2020}
{Fursyak}, Y.~A., {Kutsenko}, A.~S., \& {Abramenko}, V.~I. 2020, \solphys, 295,
  19

\bibitem[{{Gary} \& {Hagyard}(1990)}]{Gary1990}
{Gary}, G.~A., \& {Hagyard}, M.~J. 1990, \solphys, 126, 21

\bibitem[{{Holman} {et~al.}(2003){Holman}, {Sui}, {Schwartz}, \&
  {Emslie}}]{Holman2003}
{Holman}, G.~D., {Sui}, L., {Schwartz}, R.~A., \& {Emslie}, A.~G. 2003, \apjl,
  595, L97

\bibitem[{{Hurford} {et~al.}(2002){Hurford}, {Schmahl}, {Schwartz}, {Conway},
  {Aschwanden}, {Csillaghy}, {Dennis}, {Johns-Krull}, {Krucker}, {Lin},
  {McTiernan}, {Metcalf}, {Sato}, \& {Smith}}]{Hurford2002}
{Hurford}, G.~J., {Schmahl}, E.~J., {Schwartz}, R.~A., {et~al.} 2002, \solphys,
  210, 61

\bibitem[{{Joshi} {et~al.}(2018){Joshi}, {Ibrahim}, {Shanmugaraju}, \&
  {Chakrabarty}}]{Joshi2018}
{Joshi}, B., {Ibrahim}, M.~S., {Shanmugaraju}, A., \& {Chakrabarty}, D. 2018,
  \solphys, 293, 107

\bibitem[{{Joshi} {et~al.}(2016){Joshi}, {Kushwaha}, {Veronig}, \&
  {Cho}}]{Joshi2016}
{Joshi}, B., {Kushwaha}, U., {Veronig}, A.~M., \& {Cho}, K.~S. 2016, \apj, 832,
  130

\bibitem[{{Joshi} {et~al.}(2009){Joshi}, {Veronig}, {Cho}, {Bong}, {Somov},
  {Moon}, {Lee}, {Manoharan}, \& {Kim}}]{Joshi2009}
{Joshi}, B., {Veronig}, A., {Cho}, K.~S., {et~al.} 2009, \apj, 706, 1438

\bibitem[{{Kharayat} {et~al.}(2021){Kharayat}, {Joshi}, {Mitra}, {Manoharan},
  \& {Monstein}}]{Kharayat2021}
{Kharayat}, H., {Joshi}, B., {Mitra}, P.~K., {Manoharan}, P.~K., \& {Monstein},
  C. 2021, \solphys, 296, 99

\bibitem[{{Kleint} {et~al.}(2015){Kleint}, {Battaglia}, {Reardon}, {Sainz
  Dalda}, {Young}, \& {Krucker}}]{Kleint2015}
{Kleint}, L., {Battaglia}, M., {Reardon}, K., {et~al.} 2015, \apj, 806, 9

\bibitem[{{Kontogiannis} {et~al.}(2017){Kontogiannis}, {Georgoulis}, {Park}, \&
  {Guerra}}]{Kontogiannis2017}
{Kontogiannis}, I., {Georgoulis}, M.~K., {Park}, S.-H., \& {Guerra}, J.~A.
  2017, \solphys, 292, 159

\bibitem[{{Lemen} {et~al.}(2012){Lemen}, {Title}, {Akin}, {Boerner}, {Chou},
  {Drake}, {Duncan}, {Edwards}, {Friedlaender}, {Heyman}, {Hurlburt}, {Katz},
  {Kushner}, {Levay}, {Lindgren}, {Mathur}, {McFeaters}, {Mitchell}, {Rehse},
  {Schrijver}, {Springer}, {Stern}, {Tarbell}, {Wuelser}, {Wolfson}, {Yanari},
  {Bookbinder}, {Cheimets}, {Caldwell}, {Deluca}, {Gates}, {Golub}, {Park},
  {Podgorski}, {Bush}, {Scherrer}, {Gummin}, {Smith}, {Auker}, {Jerram},
  {Pool}, {Soufli}, {Windt}, {Beardsley}, {Clapp}, {Lang}, \&
  {Waltham}}]{Lemen2012}
{Lemen}, J.~R., {Title}, A.~M., {Akin}, D.~J., {et~al.} 2012, \solphys, 275, 17

\bibitem[{{Li} {et~al.}(2019){Li}, {Jaroszynski}, {Pearse}, {Orf}, \&
  {Clyne}}]{Li2019}
{Li}, S., {Jaroszynski}, S., {Pearse}, S., {Orf}, L., \& {Clyne}, J. 2019,
  Atmosphere, 10, 488

\bibitem[{{Li} {et~al.}(2015){Li}, {Ding}, {Qiu}, \& {Cheng}}]{Li2015}
{Li}, Y., {Ding}, M.~D., {Qiu}, J., \& {Cheng}, J.~X. 2015, \apj, 811, 7

\bibitem[{{Li} {et~al.}(2010){Li}, {Lynch}, {Welsch}, {Stenborg}, {Luhmann},
  {Fisher}, {Liu}, \& {Nightingale}}]{Li2010}
{Li}, Y., {Lynch}, B.~J., {Welsch}, B.~T., {et~al.} 2010, \solphys, 264, 149

\bibitem[{{Lin} {et~al.}(2002){Lin}, {Dennis}, {Hurford}, {Smith}, {Zehnder},
  {Harvey}, {Curtis}, {Pankow}, {Turin}, {Bester}, {Csillaghy}, {Lewis},
  {Madden}, {van Beek}, {Appleby}, {Raudorf}, {McTiernan}, {Ramaty}, {Schmahl},
  {Schwartz}, {Krucker}, {Abiad}, {Quinn}, {Berg}, {Hashii}, {Sterling},
  {Jackson}, {Pratt}, {Campbell}, {Malone}, {Landis}, {Barrington-Leigh},
  {Slassi-Sennou}, {Cork}, {Clark}, {Amato}, {Orwig}, {Boyle}, {Banks},
  {Shirey}, {Tolbert}, {Zarro}, {Snow}, {Thomsen}, {Henneck}, {McHedlishvili},
  {Ming}, {Fivian}, {Jordan}, {Wanner}, {Crubb}, {Preble}, {Matranga}, {Benz},
  {Hudson}, {Canfield}, {Holman}, {Crannell}, {Kosugi}, {Emslie}, {Vilmer},
  {Brown}, {Johns-Krull}, {Aschwanden}, {Metcalf}, \& {Conway}}]{Lin2002}
{Lin}, R.~P., {Dennis}, B.~R., {Hurford}, G.~J., {et~al.} 2002, \solphys, 210,
  3

\bibitem[{{Liu} {et~al.}(2015){Liu}, {Deng}, {Liu}, {Lee}, {Pariat},
  {Wiegelmann}, {Liu}, {Kleint}, \& {Wang}}]{Liu2015}
{Liu}, C., {Deng}, N., {Liu}, R., {et~al.} 2015, \apjl, 812, L19

\bibitem[{{Mitra} \& {Joshi}(2019)}]{Mitra2019}
{Mitra}, P.~K., \& {Joshi}, B. 2019, \apj, 884, 46

\bibitem[{{Mitra} \& {Joshi}(2021)}]{Mitra2021}
---. 2021, \mnras, 503, 1017

\bibitem[{{Mitra} {et~al.}(2020{\natexlab{a}}){Mitra}, {Joshi}, \&
  {Prasad}}]{Mitra2020}
{Mitra}, P.~K., {Joshi}, B., \& {Prasad}, A. 2020{\natexlab{a}}, \solphys, 295,
  29

\bibitem[{{Mitra} {et~al.}(2018){Mitra}, {Joshi}, {Prasad}, {Veronig}, \&
  {Bhattacharyya}}]{Mitra2018}
{Mitra}, P.~K., {Joshi}, B., {Prasad}, A., {Veronig}, A.~M., \&
  {Bhattacharyya}, R. 2018, \apj, 869, 69

\bibitem[{{Mitra} {et~al.}(2020{\natexlab{b}}){Mitra}, {Joshi}, {Veronig},
  {Chandra}, {Dissauer}, \& {Wiegelmann}}]{Mitra2020a}
{Mitra}, P.~K., {Joshi}, B., {Veronig}, A.~M., {et~al.} 2020{\natexlab{b}},
  \apj, 900, 23

\bibitem[{{Moore} \& {Roumeliotis}(1992)}]{Moore1992}
{Moore}, R.~L., \& {Roumeliotis}, G. 1992, {Triggering of Eruptive Flares -
  Destabilization of the Preflare Magnetic Field Configuration}, ed.
  Z.~{Svestka}, B.~V. {Jackson}, \& M.~E. {Machado}, Vol. 399, 69

\bibitem[{{Moore} {et~al.}(2001){Moore}, {Sterling}, {Hudson}, \&
  {Lemen}}]{Moore2001}
{Moore}, R.~L., {Sterling}, A.~C., {Hudson}, H.~S., \& {Lemen}, J.~R. 2001,
  \apj, 552, 833

\bibitem[{{Nitta} \& {Hudson}(2001)}]{Nitta2001}
{Nitta}, N.~V., \& {Hudson}, H.~S. 2001, \grl, 28, 3801

\bibitem[{{Pesnell} {et~al.}(2012){Pesnell}, {Thompson}, \&
  {Chamberlin}}]{Pesnell2012}
{Pesnell}, W.~D., {Thompson}, B.~J., \& {Chamberlin}, P.~C. 2012, \solphys,
  275, 3

\bibitem[{{Priest} \& {Forbes}(2002)}]{Priest2002}
{Priest}, E.~R., \& {Forbes}, T.~G. 2002, \aapr, 10, 313

\bibitem[{{R{\'e}gnier}(2012)}]{Regnier2012}
{R{\'e}gnier}, S. 2012, \solphys, 277, 131

\bibitem[{{Romano} {et~al.}(2018){Romano}, {Elmhamdi}, {Falco}, {Costa},
  {Kordi}, {Al-Trabulsy}, \& {Al-Shammari}}]{Romano2018}
{Romano}, P., {Elmhamdi}, A., {Falco}, M., {et~al.} 2018, \apjl, 852, L10

\bibitem[{{Romano} {et~al.}(2015){Romano}, {Zuccarello}, {Guglielmino},
  {Berrilli}, {Bruno}, {Carbone}, {Consolini}, {de Lauretis}, {Del Moro},
  {Elmhamdi}, {Ermolli}, {Fineschi}, {Francia}, {Kordi}, {Landi
  Degl'Innocenti}, {Laurenza}, {Lepreti}, {Marcucci}, {Pallocchia},
  {Pietropaolo}, {Romoli}, {Vecchio}, {Vellante}, \& {Villante}}]{Romano2015}
{Romano}, P., {Zuccarello}, F., {Guglielmino}, S.~L., {et~al.} 2015, \aap, 582,
  A55

\bibitem[{{Sahu} {et~al.}(2020){Sahu}, {Joshi}, {Mitra}, {Veronig}, \&
  {Yurchyshyn}}]{Sahu2020}
{Sahu}, S., {Joshi}, B., {Mitra}, P.~K., {Veronig}, A.~M., \& {Yurchyshyn}, V.
  2020, \apj, 897, 157

\bibitem[{{Sahu} {et~al.}(2022){Sahu}, {Joshi}, {Sterling}, {Mitra}, \&
  {Moore}}]{Sahu2022}
{Sahu}, S., {Joshi}, B., {Sterling}, A.~C., {Mitra}, P.~K., \& {Moore}, R.~L.
  2022, \apj, 930, 41

\bibitem[{{Schou} {et~al.}(2012){Schou}, {Scherrer}, {Bush}, {Wachter},
  {Couvidat}, {Rabello-Soares}, {Bogart}, {Hoeksema}, {Liu}, {Duvall}, {Akin},
  {Allard}, {Miles}, {Rairden}, {Shine}, {Tarbell}, {Title}, {Wolfson},
  {Elmore}, {Norton}, \& {Tomczyk}}]{Schou2012}
{Schou}, J., {Scherrer}, P.~H., {Bush}, R.~I., {et~al.} 2012, \solphys, 275,
  229

\bibitem[{{Schrijver} {et~al.}(2005){Schrijver}, {De Rosa}, {Title}, \&
  {Metcalf}}]{Schrijver2005}
{Schrijver}, C.~J., {De Rosa}, M.~L., {Title}, A.~M., \& {Metcalf}, T.~R. 2005,
  \apj, 628, 501

\bibitem[{{Shibata}(1999)}]{shibata1999}
{Shibata}, K. 1999, \apss, 264, 129

\bibitem[{{Smith} {et~al.}(2002){Smith}, {Lin}, {Turin}, {Curtis}, {Primbsch},
  {Campbell}, {Abiad}, {Schroeder}, {Cork}, {Hull}, {Landis}, {Madden},
  {Malone}, {Pehl}, {Raudorf}, {Sangsingkeow}, {Boyle}, {Banks}, {Shirey}, \&
  {Schwartz}}]{Smith2002}
{Smith}, D.~M., {Lin}, R.~P., {Turin}, P., {et~al.} 2002, \solphys, 210, 33

\bibitem[{{Tan} {et~al.}(2006){Tan}, {Ji}, {Huang}, {Zhou}, {Song}, \&
  {Huang}}]{Tan2006}
{Tan}, B., {Ji}, H., {Huang}, G., {et~al.} 2006, \solphys, 239, 137

\bibitem[{{Toriumi} \& {Wang}(2019)}]{Toriumi2019}
{Toriumi}, S., \& {Wang}, H. 2019, Living Reviews in Solar Physics, 16, 3

\bibitem[{{T{\"o}r{\"o}k} {et~al.}(2014){T{\"o}r{\"o}k}, {Leake}, {Titov},
  {Archontis}, {Miki{\'c}}, {Linton}, {Dalmasse}, {Aulanier}, \&
  {Kliem}}]{Torok2014}
{T{\"o}r{\"o}k}, T., {Leake}, J.~E., {Titov}, V.~S., {et~al.} 2014, \apjl, 782,
  L10

\bibitem[{{Vekstein}(2016)}]{Vekstein2016}
{Vekstein}, G. 2016, Journal of Plasma Physics, 82, 925820401

\bibitem[{{Vemareddy}(2017)}]{Vemareddy2017}
{Vemareddy}, P. 2017, \apj, 845, 59

\bibitem[{{Veronig} {et~al.}(2002){Veronig}, {Temmer}, {Hanslmeier}, {Otruba},
  \& {Messerotti}}]{Veronig2002}
{Veronig}, A., {Temmer}, M., {Hanslmeier}, A., {Otruba}, W., \& {Messerotti},
  M. 2002, \aap, 382, 1070

\bibitem[{{Wiegelmann}(2008)}]{Wiegelmann2008}
{Wiegelmann}, T. 2008, Journal of Geophysical Research (Space Physics), 113,
  A03S02

\bibitem[{{Wiegelmann} \& {Inhester}(2010)}]{Wiegelmann2010}
{Wiegelmann}, T., \& {Inhester}, B. 2010, \aap, 516, A107

\bibitem[{{Wiegelmann} {et~al.}(2012){Wiegelmann}, {Thalmann}, {Inhester},
  {Tadesse}, {Sun}, \& {Hoeksema}}]{Wiegelmann2012}
{Wiegelmann}, T., {Thalmann}, J.~K., {Inhester}, B., {et~al.} 2012, \solphys,
  281, 37

\bibitem[{{Woodgate} {et~al.}(1984){Woodgate}, {Martres}, {Smith}, {Strong},
  {McCabe}, {Machado}, {Gaizauskas}, {Stewart}, \& {Sturrock}}]{Woodgate1984}
{Woodgate}, B.~E., {Martres}, M.~J., {Smith}, J.~B., J., {et~al.} 1984,
  Advances in Space Research, 4, 11

\bibitem[{{Woods} {et~al.}(2017){Woods}, {Harra}, {Matthews}, {Mackay},
  {Dacie}, \& {Long}}]{Woods2017}
{Woods}, M.~M., {Harra}, L.~K., {Matthews}, S.~A., {et~al.} 2017, \solphys,
  292, 38

\bibitem[{{Woods} {et~al.}(2018){Woods}, {Inoue}, {Harra}, {Matthews},
  {Kusano}, \& {Kalmoni}}]{Woods2018}
{Woods}, M.~M., {Inoue}, S., {Harra}, L.~K., {et~al.} 2018, \apj, 860, 163

\bibitem[{{Yang} {et~al.}(2016){Yang}, {Guo}, \& {Ding}}]{Yang2016}
{Yang}, K., {Guo}, Y., \& {Ding}, M.~D. 2016, \apj, 824, 148

\bibitem[{{Young} {et~al.}(2015){Young}, {Tian}, \& {Jaeggli}}]{Young2015}
{Young}, P.~R., {Tian}, H., \& {Jaeggli}, S. 2015, \apj, 799, 218

\bibitem[{{Zhang} \& {Wang}(2002)}]{Zhang2002}
{Zhang}, J., \& {Wang}, J. 2002, \apjl, 566, L117

\bibitem[{{Zuccarello} {et~al.}(2021){Zuccarello}, {Ermolli}, {Kors{\'o}s},
  {Giorgi}, {Guglielmino}, {Erd{\'e}lyi}, \& {Romano}}]{Zuccarello2021}
{Zuccarello}, F., {Ermolli}, I., {Kors{\'o}s}, M.~B., {et~al.} 2021, Research
  in Astronomy and Astrophysics, 21, 313

\end{thebibliography}
\bibliographystyle{aasjournal}

\end{document}